\documentclass[12pt]{article}

\usepackage{amssymb}
\usepackage{amsmath}
\usepackage{amscd}
\usepackage{latexsym}
\usepackage{graphicx}

\usepackage{cite}

\newcommand{\be}{\begin{equation}}
\newcommand{\ee}{\end{equation}}
\newcommand{\Dlt}{\Delta}
\newcommand{\dlt}{\delta}
\newcommand{\prt}{\partial}
\newcommand{\br}{{\bf r}}
\newcommand{\bk}{{\bf k}}

\newcommand{\bt}{\beta}
\newcommand{\vp}{\varphi}
\newcommand{\ep}{\varepsilon}
\newcommand{\al}{\alpha}
\newcommand{\ra}{\rightarrow}
\newcommand{\sgm}{\sigma}

\newcommand{\gm}{\gamma}
\newcommand{\om}{\omega}
\newcommand{\Om}{\Omega}
\newcommand{\Gm}{\Gamma}
\newcommand{\dgr}{\dagger}
\newcommand{\lbd}{\lambda}
\newcommand{\Lbd}{\Lambda}

\newcommand{\rgl}{\rangle}
\newcommand{\lgl}{\langle}

\begin{document}

\begin{center}
{\Large {\bf Particle fluctuations in systems with Bose-Einstein condensate
} \\ [5mm]

V.I. Yukalov$^{1,2}$ } \\ [3mm]

{\it $^1$Bogolubov Laboratory of Theoretical Physics, \\
Joint Institute for Nuclear Research, Dubna 141980, Russia \\ [2mm]

$^2$Instituto de Fisica de S\~ao Carlos, Universidade de S\~ao Paulo, \\
CP 369, S\~ao Carlos 13560-970, S\~ao Paulo, Brazil} \\ [3mm]

{\bf E-mail}: yukalov@theor.jinr.ru

\end{center}

\vskip 2cm

\begin{abstract}

Particle fluctuations in systems, exhibiting  Bose-Einstein condensation, are reviewed
in order to clarify the basic points that attract high interest and often confront 
misunderstanding. It is explained that the so-called ``grand canonical catastrophe",
claiming the occurrence of catastrophic particle fluctuations in the condensed phase, 
treated by grand canonical ensemble, does not exist. What exists is the incorrect 
use of the grand canonical ensemble, where gauge symmetry is not broken, while the
correct description of the condensed phase necessarily requires gauge symmetry 
breaking. The ideal Bose gas has no catastrophic condensate fluctuations, and moreover 
there are no condensate fluctuations at all, as soon as gauge symmetry is broken. 
However it does have anomalous fluctuations of uncondensed particles, which implies 
its instability. For interacting particles, there are no condensate fluctuations, 
as soon as gauge symmetry is broken, and anomalous fluctuations of uncondensed 
particles, when correctly calculated, do not appear. Particle fluctuations in the 
systems of trapped atoms are discussed. Canonical ensemble and grand canonical 
ensemble with broken gauge symmetry are equivalent with respect to the number of 
particle scaling. 
 
\end{abstract}

\newpage

\section*{Content}

{\parindent=0pt
{\bf 1}. Introduction

\vskip 3mm
{\bf 2}. Thermodynamic stability of statistical systems

\vskip 3mm
{\bf 3}. Isothermal compressibility and particle fluctuations

\vskip 3mm
{\bf 4}. General criterion of condensate existence

\vskip 3mm
{\bf 5}. Gauge symmetry breaking and condensation

\vskip 2mm
\hspace{0.4cm}    5.1. Method of quasi-averages
\vskip 2mm
\hspace{0.4cm}    5.2. Condition for condensation
\vskip 2mm
\hspace{0.4cm}    5.3. Ginibre theorem
\vskip 2mm
\hspace{0.4cm}    5.4. Bogolubov theorem
\vskip 2mm
\hspace{0.4cm}    5.5. Roepstorff theorem
\vskip 2mm
\hspace{0.4cm}    5.6. Bogolubov shift

\vskip 3mm
{\bf 6}. Spurious catastrophic condensate fluctuations

\vskip 3mm
{\bf 7}. There is no any ``grand canonical catastrophe"

\vskip 3mm
{\bf 8}. Ideal uniform Bose-condensed gas is unstable

\vskip 3mm
{\bf 9}. Bose-Einstein condensation in interacting systems

\vskip 3mm
{\bf 10}. Particle fluctuations in interacting systems

\vskip 3mm
{\bf 11}. Ideal gas in a rectangular box

\vskip 2mm
\hspace{0.5cm}    11.1. Condensation temperature in a box
\vskip 2mm
\hspace{0.5cm}    11.2. Fluctuations in normal gas $(T > T_c)$
\vskip 2mm
\hspace{0.5cm}    11.3. Fluctuations in condensed gas $(T < T_c)$

\vskip 3mm
{\bf 12}. Ideal gas in a power-law potential

\vskip 2mm
\hspace{0.5cm}    12.1. Condensation temperature in a power-law potential
\vskip 2mm
\hspace{0.5cm}    12.2. Fluctuations above condensation temperature $(T > T_c)$
\vskip 2mm
\hspace{0.5cm}    12.3. Fluctuations below condensation temperature $(T < T_c)$

\vskip 3mm
{\bf 13}. Fluctuations of composite observables

\vskip 3mm
{\bf 14}. Particle fluctuations and ensemble equivalence

\vskip 3mm
{\bf 15}. Conclusion
}

\section{Introduction}

The study of particle fluctuations in systems with Bose-Einstein condensate is of 
high importance, since these fluctuations characterize the system stability, hence 
the ability of realizing experiments with these systems, the competence for providing
correct interpretations of the experiments, and the possibility of employing the 
obtained knowledge for different practical applications.  

Systems with Bose-Einstein condensate are nowadays widely spread, including, in addition 
to liquid helium \cite{London_1,Khalatnikov_2,Putterman_3,Tilley_4}, condensates of 
trapped atoms bound in traps of various geometry \cite{Parkins_5,Dalfovo_6,Courteille_7,
Andersen_8,Yukalov_9,Bongs_10,Yukalov_11,Posazhennikova_12,Yukalov_13,Proukakis_14,
Yurovsky_15,Yukalov_16,Yukalov_17,Yukalov_18} and in optical lattices \cite{Morsch_19,
Moseley_20,Yukalov_21,Krutitsky_22,Yukalov_23}, pion condensate \cite{Migdal_24}, and 
condensates of six-quark and, possibly, other quark clusters \cite{Yukalov_25,Yukalov_26}. 
Some coherent nonequilibrium processes can be interpreted as nonequilibrium condensation. 
For example, coherent procession of spins can be treated as magnon condensation 
\cite{Demokritov_27,Bunkov_28}. Equilibrium quasiparticles, whose number is not conserved,
are not able to form Bose condensates \cite{Yukalov_29}. But by means of external pumping, 
supporting the density of the qusiparticles, sufficient for their condensation, one can 
create quasistationary steady-states with nonequilibrium condensates having much in 
common with equilibrium counterparts. Examples of such nonequilibrium quasistationary 
steady-state condensates are magnon condensates \cite{Demokritov_27,Bunkov_28}, exciton 
condensates \cite{Yoshioka_30}, and photon condensate \cite{Klaers_31,Wang_48}.

Here, we do not touch nonequilibrium pumped condensates, but shall concentrate on 
equilibrium condensates, since even in their theory there is a number of delusions, 
especially with regard to particle fluctuations. Before passing to nonequilibrium cases, 
it is necessary to clean out the fallacies related to the standard equilibrium 
condensates. First of all, beginning to study the properties of a system, it is 
necessary to check wether this system can in principle exist as a stable equilibrium 
object.  

Below, the system of units is accepted, where the Boltzmann constant $k_B$ and Planck
constant $\hbar$ are set to unity.

\section{Thermodynamic stability of statistical systems}

Statistical systems comprise a large number of particles $N \gg 1$ that can be finite or
asymptotically large, when $N \ra \infty$. The situation where the number of particles 
$N \ra \infty$ and the system volume $V \ra \infty$ is named the {\it thermodynamic limit}, 
\be
\label{1}
 N \ra \infty \; , \qquad V \ra \infty \; , \qquad \frac{N}{V} \ra const \;   .
\ee
Thermodynamic and observable quantities are classified as extensive and intensive.

{\it Extensive} quantities, say $A_N$, for a system of $N$ particles, in the 
thermodynamic limit behave as
\be
\label{2}
N \ra \infty \; , \qquad A_N \ra \infty \; , \qquad \frac{A_N}{N} \ra const \; ,
\ee
which is a generalized form of the thermodynamic limit that is valid both for infinitely 
large systems, as well as for systems confined by means of external potentials or walls  
\cite{Yukalov_32}. 

{\it Intensive} quantities can be represented as the ratio of extensive quantities, so 
that such a ratio, say $A_N/N$, is finite for all $N$, including $N \ra \infty$, 
\be
\label{3}
\frac{|\; A_N\; |}{N} \; < \; \infty \qquad (\forall N ) \;   .
\ee

An {\it equilibrium state} of a statistical system corresponds to an extremum of a 
thermodynamic potential with respect to the variation of parameters. This, depending on 
the chosen thermodynamic variables, can be internal energy $E$, free energy $F$, Gibbs
potential $G$, or grand potential $\Omega$,
\be
\label{4}
\dlt E \; = \; \dlt F \; = \; \dlt G \; = \; \dlt \Om \; = \; 0 \;  .
\ee

The extremum is a minimum under the {\it thermodynamic stability condition}
\be
\label{5}
\dlt^2 E \; > \; 0 \; , \qquad \dlt^2 F \; > \; 0 \; , \qquad \dlt^2 G \; > \; 0 \; , 
\qquad \dlt^2 \Om \; > \; 0 \;   .
\ee
If the stability condition does not hold, the extremum is related to an unstable state. 
A local minimum describes a metastable state. The absolute minimum defines the absolutely
stable equilibrium state.  

The stability condition (\ref{5}) results in the requirement of positivity (or at least
semi-positivity) for susceptibilities, such as specific heat, magnetic susceptibility,
and isothermic compressibility. Of the major interest here is the susceptibility 
connected with particle fluctuations, which is the isothermic compressibility that can 
be defined through the derivatives
\be
\label{6}
\varkappa_T \; = \; 
- \; \frac{1}{V} \; \left( \frac{\prt V}{\prt P} \right)_{TN} \; = \;
 - \; \frac{1}{V} \; \left( \frac{\prt P}{\prt V} \right)^{-1}_{TN} \;  ,
\ee
in which temperature $T$ and the number of particles $N$ are fixed; $P$ is pressure
and $V$ is the system volume. Using the notation for the density $\rho \equiv N/V$, 
it is straightforward to write
\be
\label{7}
 \varkappa_T \; = \; 
\frac{1}{\rho} \; \left( \frac{\prt\rho}{\prt P}\right)_{TN} \; = \;
\frac{1}{\rho} \; \left( \frac{\prt P}{\prt\rho}\right)_{TN}^{-1}  \; .
\ee
This is equivalent to the derivatives 
\be
\label{8}
\varkappa_T \; = \; 
-\; \frac{1}{V} \; \left( \frac{\prt^2 G}{\prt P^2}\right)_{TN} \; = \; 
\frac{1}{V} \; \left( \frac{\prt^2 F}{\prt V^2}\right)_{TN}^{-1}  \;   .
\ee
When temperature $T$ and volume $V$ are fixed, then the compressibility is
\be
\label{9}
 \varkappa_T \; = \; 
\frac{V}{\rho^2} \; \left( \frac{\prt^2 F}{\prt \rho^2}\right)_{TV}^{-1} \; =  \;
-\; \frac{1}{\rho N} \; \left( \frac{\prt^2 \Om}{\prt \mu^2}\right)_{TV}  \;   ,
\ee
which is the same as 
\be
\label{10}
 \varkappa_T \; = \; 
\frac{1}{\rho N} \; \left( \frac{\prt N}{\prt \mu}\right)_{TV} \; =  \;
\frac{1}{N} \; \left( \frac{\prt N}{\prt P}\right)_{TV} \;   ,
\ee
where $\mu$ is chemical potential. 

The isothermic compressibility, as prescribed by the stability condition (\ref{5}), has 
to be non-negative and, being an intensive observable quantity, it has to be finite
according to definition (\ref{3}), hence
\be
\label{11}
0 \; \leq \; \varkappa_T \; < \; \infty \qquad (\forall N)
\ee
for all $N$ including $N \ra \infty$. Let us emphasize that the discussed above 
finiteness concerns variations with respect to the number of particles, under fixed 
other thermodynamical parameters. As a function of some other thermodynamic variables, 
the compressibility can diverge at the points of phase transitions, where the system 
becomes unstable.    

Remembering that the sound velocity squared is given by the expression
\be
\label{12}
 s^2 \; = \; \frac{1}{m} \; \left( \frac{\prt P}{\prt \rho}\right)_{TN} \;  ,
\ee
we see that the sound velocity and compressibility are mutually related,
\be
\label{13}
 \varkappa_T \; = \; \frac{1}{m \rho s^2} \; , \qquad 
s^2 \; = \; \frac{1}{m \rho \varkappa_T} \;  .
\ee

The above formulas are general and exact thermodynamic relations that are valid 
for all equilibrium stable statistical systems. The stability condition (\ref{11})
is compulsory for any stable equilibrium system. If this condition is broken, this 
means that either the considered system is not equilibrium and stable or calculations 
are incorrect.

\section{Isothermal compressibility and particle fluctuations}

From expression (\ref{10}) it follows that the isothermal compressibility can be 
considered as a measure of the strength of particle fluctuations since it is 
characterized by the number-of-particle variance
\be
\label{14}
 \varkappa_T \; = \; \frac{{\rm var}(\hat N)}{\rho T N} \;  ,
\ee
in which the variance is 
\be
\label{15}
  {\rm var}(\hat N) \; \equiv \; \lgl \; \hat N^2 \; \rgl - 
\lgl \; \hat N \; \rgl^2 \; ,
\ee
with the number-of-particle operator in the second-quantization representation
\be
\label{16} 
\hat N \; = \; \int \psi^\dgr(\br) \; \psi(\br) \; d\br \;   .
\ee
Here $\psi({\bf r})$ is the field operator satisfying Bose commutation relations and the 
angle brackets denote statistical averaging with a statistical operator $\hat{\rho}$.
The spatial coordinates ${\bf r}$ are shown explicitly. If there are some internal 
variables, such as spin or component labels, the field operators can be treated as columns 
in these variables. 

In its turn, variance (\ref{15}) can be expressed as
\be
\label{17}
{\rm var}(\hat N) \; = \; N + 
\int \rho(\br) \; \rho(\br') \; [ \; g(\br,\br') - 1 \; ] \; d\br d\br' 
\ee
through the pair correlation function
\be
\label{18}
  g(\br,\br') \;  = \; \frac{\rho_2(\br,\br',\br,\br')}{\rho(\br)\rho(\br')} \;  ,
\ee
where the notations for the particle density
\be
\label{19}
\rho(\br) \; \equiv \; \lgl \; \psi^\dgr(\br) \; \psi(\br) \; \rgl
\ee
and the second-order reduced density matrix
$$
\rho_2(\br,\br',\br,\br') \; = \; {\rm Tr} \; \psi(\br) \; \psi(\br') \; \hat\rho \;
\psi^\dgr(\br') \; \psi^\dgr(\br) \; =
$$
\be
\label{20}
= \; \lgl \; \psi^\dgr(\br') \; \psi^\dgr(\br)
\; \psi(\br) \; \psi(\br') \; \rgl = 
\lgl \; \psi^\dgr(\br) \; \psi^\dgr(\br')
\; \psi(\br') \; \psi(\br) \; \rgl
\ee
are used. More information on reduced density matrices can be found in Ref. 
\cite{Coleman_33}.  

The pair correlation function is connected with the static structure factor
\be
\label{21}
 S(\bk) \; = \; 
1 + \frac{1}{N} \int \rho(\br) \; \rho(\br') \; [ \; g(\br,\br') - 1 \; ] \;
e^{-i\bk \cdot(\br-\br')} \;  d\br d\br' \;   ,
\ee
whose long-wave limit is
\be
\label{22}
 S(0) \; = \; 1 + \frac{1}{N} \int \rho(\br) \; \rho(\br') \; [ \; g(\br,\br') - 1 \; ] \;
 d\br d\br' \;  .
\ee

In the particular case of a uniform system, where
\be
\label{23}
\rho(\br) \; = \; \rho \; , \qquad  g(\br,\br') = g(\br-\br') \;  ,
\ee
the structure factor takes the form
\be
\label{24}
S(\bk) \; = \; 1 + \rho \int [ \; g(\br) - 1 \; ] \; e^{-i\bk\cdot\br}\; d\br \; ,
\ee
and its limit reads as
\be
\label{25}
 S(0) \; = \; 1 + \rho \int [ \; g(\br) - 1 \; ] \; d\br \; .
\ee
Since the structure factor can be measured, the pair correlation function is also 
measurable,
\be
\label{26}
 g(\br) \; = \; 1 + \frac{1}{\rho} \int [\; S(\bk) - 1 \; ]  e^{i\bk\cdot\br}\; 
\frac{d\bk}{(2\pi)^3} \; .
\ee

In that way, particle fluctuations of a uniform system are characterized by the variance
\be
\label{27}
 {\rm var}(\hat N) \; = \; N + \rho N \int [\; g(\br) - 1 \; ] \; d\br \;  .
\ee
As far as the structure factor is measurable, particle fluctuations can be measured by
studying the structure factor or the pair correlation function and using the relations
\be
\label{28}
 \frac{{\rm var}(\hat N)}{N} \; = \; S(0) \; = \; \frac{T}{m s^2} \; = \; 
\rho T \varkappa_T \;  .
\ee
The pair correlation function is assumed to obey the boundary condition
\be
\label{29}
\lim_{|\;\br\;|\ra \infty} \; g(\br) \; = \; 1 \;   .
\ee 

For a nonuniform system, the number-of-particle variance takes the form 
\cite{Yukalov_34,Yukalov_35}
\be
\label{30}
 {\rm var}(\hat N) \; = \; \frac{T}{m} \int \frac{\rho(\br)}{s^2(\br)} \; d\br \;  .
\ee

In any case, due to the stability requirement (\ref{11}) for the isothermal compressibility, 
in a stable equilibrium system, it is necessary that particle fluctuations satisfy the 
stability condition 
\be
\label{31}
  0 \; \leq \; \frac{  {\rm var}(\hat N) }{N} \; < \; \infty
\ee
for all $N$ including $N \ra \infty$.

Again it is important to stress that if the stability condition (\ref{31}) is not valid,
then either the considered system is not in a state of stable equilibrium or something 
is wrong with calculations.

\section{General criterion of condensate existence} 

The above relations are valid for any stable equilibrium statistical system. To specify 
the consideration to systems with Bose-Einstein condensate, it is necessary to concretize
the description. 

First, we need to define a basis of functions in order to realize expansions of field 
operators. Among all admissible functions there is a natural preferable set reflecting 
the underlying physics. This is a set of eigenfunctions of the reduced density matrix
\be
\label{32}
\rho(\br,\br') \; = \; {\rm Tr} \; \psi(\br) \; \hat\rho \; \psi^\dgr(\br') \; = \;
\lgl \; \psi^\dgr(\br') \; \psi(\br) \; \rgl
\ee
defined by the eigenproblem
\be
\label{33}
 \int \rho(\br,\br') \; \vp_k(\br') \; d\br' \; = \; n_k \vp_k(\br) \;  ,
\ee
where $k$ is a quantum multi-index labelling the eigenfunctions, which are termed 
\cite{Coleman_33} {\it natural orbitals}. The eigenvalues $n_k$ play the role of the
occupation numbers for the states labelled by the index $k$. From the expression
\be
\label{34}
 n_k \; = \; \int  \vp_k(\br) \; \rho(\br,\br') \; \vp_k(\br') \; d\br  d\br' \; ,
\ee
assuming that the basis of the natural orbitals is orthonormal, it follows the 
normalization
\be
\label{35}
 \sum_k n_k \; = \; \int \rho(\br) \; d\br \; = \; N \;  ,
\ee
since
\be
\label{36}
\rho(\br) \; = \; \rho(\br,\br) \;  .
\ee
 
The largest eigenvalue
\be
\label{37}
N_0 \; \equiv \; \sup_k n_k
\ee
is associated with the number of condensed particles \cite{Penrose_36}. The Bose-Eistein
condensate in a stable statistical system exists when the number of condensed particles 
$N_0$ is macroscopic,
\be
\label{38}
 \lim_{N\ra\infty} \; \frac{N_0}{N} \; > \; 0 \;  .
\ee
Here the limit $N \ra \infty$ implies the thermodynamic limit.   

Expanding the field operators over the natural orbitals, it is possible to separate the 
parts $\psi_0$, related to the condensate, and $\psi_1$, corresponding to uncondensed
particles,
\be
\label{39}
 \psi(\br) \; = \; \sum_k a_k \; \vp_k(\br) \; = \; \psi_0(\br) + \psi_1(\br) \;  ,
\ee
where the field operators of condensed and uncondensed particles are
\be
\label{40}
 \psi_0(\br) \; = \; a_0 \vp_0(\br) \; , \qquad  
\psi_1(\br) \; = \; \sum_{k\neq 0} a_k \; \vp_k(\br) \; .
\ee
By this definition, one has
\be
\label{41}
\int \psi_0^\dgr(\br) \; \psi_1(\br) \; d\br \; = \; 0
\ee
due to the orthogonality of natural orbitals.

Assuming the quantum-number conservation condition
\be
\label{42}
  \lgl \; a_k^\dgr \; a_p \; \rgl \; = \; 
\dlt_{kp} \; \lgl \; a_k^\dgr \; a_k \; \rgl
\ee
yields
\be
\label{43}
 \lgl \;  \psi_0^\dgr(\br) \; \psi_1(\br) \; \rgl \; = \; 0 \;  .
\ee

The operator for the number of condensed particles is
\be
\label{44}
\hat N_0 \; = \; \int \psi_0^\dgr(\br) \; \psi_0(\br) \; d\br \; = \; 
a_0^\dgr \; a_0
\ee
and the operator for the number of uncondensed particles is
\be
\label{45}
\hat N_1 \; = \; \int \psi_1^\dgr(\br) \; \psi_1(\br) \; d\br \; = \;
\sum_{k\neq 0} a_k^\dgr \; a_k  \; .
\ee
This gives the operator for the total number of particles
\be
\label{46}
  \hat N \; = \; \hat N_0 + \hat N_1 \; .
\ee

The number of condensed particles is the average
\be
\label{47}
N_0 \; = \; \lgl \;  \hat N_0 \; \rgl \; = \; \lgl \; a_0^\dgr \; a_0 \; \rgl 
\ee
and the number of uncondensed particles is
\be
\label{48}
N_1 \; = \; \int \lgl \; \psi_1^\dgr(\br) \; \psi_1(\br) \; \rgl \; d\br \; = \;
\sum_{k\neq 0} \lgl \; a_k^\dgr \; a_k \; \rgl \;  .
\ee

The criterion of condensate existence (\ref{38}) can be written in the form
\be
\label{49}
 \lim_{N\ra\infty} \; \frac{\lgl \; a_0^\dgr \; a_0 \; \rgl}{N} \; > \; 0 \;  .
\ee

\section{Gauge symmetry breaking and condensation}

The standard Hamiltonian $H_N$ for a system of $N$ particles with binary interactions
enjoys the global gauge symmetry $U(1)$, so that the Hamiltonian is invariant under 
the replacement
\be
\label{50}
 \psi(\br) \; \longmapsto \; \psi(\br) e^{i\al} \;  ,
\ee
where $\alpha$ is a real number. Bose-Einstein condensation is associated with the 
global gauge symmetry breaking that can be realized in several ways 
\cite{Yukalov_37,Yukalov_38}. Sometimes one claims that the gauge symmetry breaking 
is just a convenient technical tool, but Bose condensation does not require the gauge
symmetry to be broken. Below, this problem is thoroughly analyzed.

\subsection{Method of quasi-averages} 

A very instructive method of symmetry breaking is the Bogolubov method of 
quasi-averages \cite{Bogolubov_39,Bogolubov_40,Bogolubov_41}. According to this 
method, the symmetry invariant Hamiltonian $H_N$ is supplemented with a symmetry 
breaking term ${\hat B}_N$, so that the Hamiltonian
\be
\label{51} 
H_\ep \; = \; H_N + \ep \hat B_N
\ee
becomes not invariant with respect to the initial symmetry. Here $\varepsilon$ is a 
real parameter and the term $\hat{B}_N$ breaks the gauge symmetry.

The quasi-average for an operator $\hat{A}_N$ is defined through the double limit
\be
\label{52}
 \lim_{N\ra\infty} \; \frac{\lgl \; \hat A_N \; \rgl}{N} \; \equiv \;  
\lim_{\ep\ra 0} \; \lim_{N\ra\infty} \; \frac{\lgl \; \hat A_N \; \rgl_\ep}{N} \; ,
\ee
where the averaging is taken with the Hamiltonian (\ref{51}) and the limit 
$\varepsilon \ra 0$ has to be taken necessarily after the thermodynamic limit 
$N \ra \infty$. 

In principle, it is admissible, instead of a fixed $\varepsilon$, to accept 
$\varepsilon \propto 1/N^\alpha$ with $0 < \alpha < 1$, so that it is sufficient 
to take just a single thermodynamic limit \cite{Yukalov_51,Yukalov_52,Coleman_53}.
However, for simplicity, in what follows we use the standard method of quasi-averages.    

Spontaneous gauge symmetry breaking implies that
\be
\label{53}
\lim_{\ep\ra 0} \; \lim_{N\ra\infty} \; 
\frac{1}{N} \int |\; \lgl \; \psi_0(\br) \; \rgl_\ep \; |^2 \; d\br \; > \; 0 \;  ,
\ee
which is equivalent to
\be
\label{54}
\lim_{\ep\ra 0} \; \lim_{N\ra\infty} \; 
\frac{|\; \lgl \; a_0 \; \rgl_\ep\; |^2}{N} \; > \; 0 \;  .
\ee

There are several rigorously proved mathematical facts connecting the spontaneous 
breaking of global gauge symmetry with Bose-Einstein condensation.

\subsection {Condition for Condensation}
 
By resorting to the Cauchy-Schwarz inequality 
$|\langle a_0 \rangle_\varepsilon|^2  \leq \langle a_0^\dgr a_0 \rangle_\varepsilon$, 
it is easy to show that from the condition (\ref{54}) of spontaneous symmetry breaking 
the condensate existence follows:
\be
\label{55}
 \lim_{\ep\ra 0} \; \lim_{N\ra\infty} \; 
\frac{\lgl \; a_0^\dgr \; a_0 \; \rgl_\ep}{N} \; > \; 0 \;   ,
\ee
hence
\be
\label{56}
 \lim_{\ep\ra 0} \; \lim_{N\ra\infty} \; 
\frac{\lgl \; \hat N_0 \; \rgl_\ep}{N} \; > \; 0 \;  .
\ee

\subsection{Ginibre theorem}

Using representation (\ref{39}) makes the system Hamiltonian (\ref{51}) a functional 
of $\psi_0$ and $\psi_1$,
\be
\label{57}
H_\ep \; = \; H_\ep[\psi_0,\; \psi_1] \;   .
\ee
In equilibrium, the grand thermodynamic potential writes as
\be
\label{58}
\Om_\ep [\psi_0,\; \psi_1] \; = \; 
- T \ln {\rm Tr} \; \exp\left\{ - \bt  H_\ep[\psi_0,\; \psi_1] \right\} \;  ,
\ee
with $\beta \equiv 1/T$ being the inverse temperature. Replacing the operator $\psi_0$
by a nonoperator quantity $\eta$ gives the thermodynamic potential
\be
\label{59}  
\Om_\ep [\eta,\; \psi_1] \; = \; 
- T \ln {\rm Tr} \; \exp\left\{ - \bt  H_\ep[\eta,\; \psi_1] \right\} \;   ,
\ee
with $\eta$ assumed to be a minimizer of the thermodynamic potential,
\be
\label{60}
 \Om_\ep [\eta,\; \psi_1] \; = \; \min_{\eta'}  \Om_\ep [\eta',\; \psi_1] \;  .
\ee

The Ginibre theorem states \cite{Ginibre_42} that in the thermodynamic limit the 
thermodynamic potentials (\ref{58}) and (\ref{59}) coincide:
\be
\label{61}
\lim_{N\ra\infty} \; \frac{1}{N} \; \Om_\ep [\psi_0,\; \psi_1] \; = \;
\lim_{N\ra\infty} \; \frac{1}{N} \;  \Om_\ep [\eta,\; \psi_1] \;  .
\ee

\subsection{Bogolubov theorem}

Let us consider correlation functions $C_\varepsilon[\psi_0,\psi_1]$ representing the 
averages of the products of the field operators. Replacing the operator $\psi_0$ by
a nonoperator quantity $\eta$ gives the correlation function $C_\varepsilon[\eta,\psi_1]$.
Bogolubov has showed \cite{Bogolubov_39,Bogolubov_40} that in the thermodynamic limit 
these correlation functions coincide:
\be
\label{62}
\lim_{N\ra\infty} C_\ep [\psi_0,\; \psi_1] \; = \;
\lim_{N\ra\infty} C_\ep [\eta,\; \psi_1] \;   .
\ee
In particular,
\be
\label{63}
\lim_{N\ra\infty} \lgl \; \psi_0(\br)\; \rgl_\ep \; = \; \eta(\br) \; , 
\qquad
\lim_{N\ra\infty} \lgl \; \psi_0^\dgr(\br)\; \psi_0(\br) \; \rgl \; = \; 
|\; \eta(\br) \; |^2 \; .
\ee

\subsection{Roepstorff theorem}

Roepstorff \cite{Roepstorff_43} has proved (see also \cite{Lieb_44,Suto_45,Lieb_46})
that Bose-Einstein condensation implies spontaneous gauge symmetry breaking:
\be
\label{64}
\lim_{N\ra\infty} \; \frac{\lgl \; a_0^\dgr \; a_0 \; \rgl}{N} \; \leq \; 
\lim_{\ep\ra 0} \; \lim_{N\ra\infty} \;
\frac{|\; \lgl \; a_0 \; \rgl_\ep \; |^2}{N} \; .
\ee
Here the left-hand side is taken without involving gauge symmetry breaking.

\subsection{Bogolubov shift}

The above theorems allow for the formulation of a general and convenient method for 
gauge symmetry breaking which is equivalent to the method of quasi-averages. For this
purpose, the field-operator representation as the sum (\ref{39}) can be written in 
the form of the {\it Bogolubov shift}
\be
\label{65}
 \psi(\br) \; = \; \eta(\br) + \psi_1(\br) \;  ,
\ee
in which $\eta({\bf r})$ is the condensate function and $\psi_1({\bf r})$ is the field
operator for uncondensed particles. The condensate function plays the role of an order
parameter satisfying the equation
\be
\label{66}
 \eta(\br) \; = \; \lgl \; \psi(\br) \; \rgl  
\ee  
breaking the gauge symmetry and defining the number of condensed particles
\be
\label{67}
N_0 \; = \; \int |\; \eta(\br) \; |^2 \; d\br \;   .
\ee
Respectively, for the field operator of uncondensed particles it follows
\be
\label{68}
 \lgl \; \psi_1(\br) \; \rgl \; = \; 0 \;  .
\ee
Similarly to condition (\ref{41}), one has the condition of orthogonality
\be
\label{69}
 \int \eta^*(\br) \; \psi_1(\br) \; d\br \; = \; 0 \;  .
\ee

It is important to emphasize that the Bogolubov shift (\ref{65}) is not an approximation, 
as one sometimes writes, but a rigorous canonical transformation valid for any number
of condensed and uncondensed particles \cite{Yukalov_47}. More discussions and 
explanations can be found in the review articles \cite{Yukalov_9,Yukalov_13,Yukalov_16,
Yukalov_17}. The above theorems are accurate mathematical results leading to the following 
conclusion.

\vskip 2mm

{\bf Conclusion}: {\it Spontaneous gauge symmetry breaking is the necessary and 
sufficient condition for Bose–Einstein condensation}.

\section{Spurious catastrophic condensate fluctuations} 

It has been noticed long time ago \cite{Fujiwara_49,Ziff_50} that the ideal Bose gas 
below the condensation temperature can exhibit catastrophically large condensate 
fluctuations, if the grand canonical ensemble is straightforwardly employed. Then the 
variance of the condensate number of particles is proportional to $N_0^2$. At the same 
time, the canonical ensemble does not show such catastrophic condensate fluctuations.
One often even claims that the grand canonical ensemble cannot be used for describing
Bose-condensed systems, since the appearance of catastrophic fluctuations signifies 
the system instability. It is useful to recall how such catastrophic particle 
fluctuations appear in the process of calculations. 

Actually, the blind use of the grand canonical ensemble leads to catastrophic
condensate fluctuations not merely for the ideal Bose gas but for any interacting Bose
system with condensate. This can be shown as follows.

Let us consider the variance
\be
\label{70}
 {\rm var}(\hat N_0) \; = \; \lgl \; \hat N_0^2 \; \rgl - \lgl \; \hat N_0\; \rgl^2
\ee
of the condensate number-of-particle operator
\be
\label{71}
 \hat N_0 \; = \; a_0^\dgr \; a_0 \;  .
\ee
The first term in (\ref{70}) is
\be
\label{72}
 \lgl \; \hat N_0^2 \; \rgl \; = \; 
\lgl \; a_0^\dgr \; a_0 \; a_0^\dgr \; a_0 \; \rgl \; = \; 
\lgl \; a_0^\dgr \; a_0^\dgr \; a_0 \; a_0 \; \rgl + 
 \lgl \; a_0^\dgr \; a_0 \; \rgl \; .
\ee
With the help of the Wick theorem, one has
$$
\lgl \; a_0^\dgr \; a_0^\dgr \; a_0 \; a_0 \; \rgl \; = \; 2 N_0^2 \;  ,
$$
so that one gets
$$
\lgl \; \hat N_0^2  \; \rgl \; = \; 2N_0^2 + N_0 \; .
$$
Then the variance (\ref{70}) reads as
\be
\label{73}
 {\rm var}(\hat N_0) \; = \; N_0 ( 1 + N_0 ) \;   ,
\ee
independently of whether the gas is ideal or not.

The existence of Bose-Einstein condensate presupposes that the number of condensed 
particles is macroscopic in the sense of criterion (\ref{38}), hence $N_0 \propto N$.   
Then the variance (\ref{73}) is proportional to $N^2$. Therefore the variance of the 
total number of particles
\be
\label{74}
{\rm var}(\hat N) \; = \; {\rm var}(\hat N_0) + {\rm var}(\hat N_1) 
\ee
is also of the order of $N^2$. As a result, the compressibility of the system 
\be
\label{75}
\varkappa_T \; = \; \frac{{\rm var}(\hat N)}{\rho T N} \; \simeq \; 
\frac{N_0(1+N_0)}{\rho T N}
\ee
diverges as
\be
\label{76}
 \varkappa_T \; \propto \; N \qquad ( N_0 \sim N ) \;  ,
\ee
which contradicts the conditions of stability (\ref{11}) and (\ref{31}) explained 
in sections 2 and 3. This is why the condensate fluctuations, characterized by the 
variance (\ref{73}) are classified as catastrophic and their appearance is named the
``grand canonical catastrophe".

\section{There is no any ``grand canonical catastrophe"}

The reason for the occurrence of catastrophic condensate fluctuations, when blindly
applying the grand canonical ensemble, is very simple \cite{Haar_54}: One forgets 
that, as explained in Sec. 5, for the existence of the Bose-Einstein condensate it 
is necessary and sufficient that gauge symmetry be broken. Without breaking the gauge 
symmetry, no condensate is possible, hence $N_0$ cannot be macroscopic, being of the 
order of $N$. So, there is no any ``grand canonical catastrophe", but there are incorrect 
calculations. Below it is shown how the condensate fluctuations are to be correctly 
treated in the frame of the grand canonical ensemble. 

Let us consider the ideal Bose gas with the Hamiltonian
\be
\label{77}
 H_N \; = \; \sum_k \left( \frac{k^2}{2m} \; - \; \mu\right) \; a_k^\dgr \; a_k \;  .
\ee
This Hamiltonian is gauge invariant under the transformation 
$a_k \mapsto a_k e^{i \varphi}$, where $\varphi$ is a real number. To break the gauge 
symmetry, it is possible to follow the method of quasi-averages, either in the form 
of thermodynamic quasi-averages \cite{Yukalov_52,Coleman_53} or in the standard 
Bogolubov form \cite{Bogolubov_39,Bogolubov_40,Bogolubov_41}. Below, the Bogolubov 
form is used. To this end, we define the Hamiltonian
\be
\label{78}
 H_\ep \; = \; H_N + \ep \hat B_N \; ,
\ee
with the symmetry breaking term
\be
\label{79}
\hat B_N  \; = \; \sqrt{V} \; \left(   a_0^\dgr e^{i\vp} + a_0 e^{-i\vp} \right) \; ,
\ee
in which $\varphi$ is fixed. Explicitly, the Hamiltonian becomes
\be
\label{80}
H_\ep \; = \; \sum_{k\neq 0} \left( \frac{k^2}{2m} \; - \; \mu\right) \; a_k^\dgr \; a_k -
\mu \; a_0^\dgr \; a_0 + 
\ep \; \sqrt{V}  \; \left( a_0^\dgr e^{i\vp} + a_0 e^{-i\vp} \right) \;  .
\ee

The Hamiltonian can be diagonalized (see e.g. \cite{Yukalov_55}) by means of the 
canonical transformation 
\be
\label{81}
a_0 \; = \; b_0 + \ep \; \frac{\sqrt{V}}{\mu} \; e^{i\vp}
\ee
resulting in the Hamiltonian
\be
\label{82}
H_\ep \; = \; \sum_{k\neq 0} \left( \frac{k^2}{2m} \; - \; \mu\right) \; a_k^\dgr \; a_k -
\mu \; b_0^\dgr \; b_0 + \frac{\ep^2}{\mu} \; V \;  .
\ee

With the diagonal Hamiltonian (\ref{82}), it is easy to find the averages
\be
\label{83}
\lgl \; b_0^\dgr \; b_0 \; \rgl_\ep \; = \; \frac{1}{e^{-\bt\mu}-1} \; = \; \frac{z}{1-z} \; , 
\qquad
\lgl \; b_0 \; \rgl_\ep \; = \; 0  \qquad \left( z \equiv e^{\bt\mu}\right) \; ,
\ee
from where one gets
\be
\label{84}
\lgl \; a_0^\dgr \; a_0 \; \rgl_\ep \; = \; \lgl \; b_0^\dgr \; b_0 \; \rgl +
\frac{\ep^2}{\mu^2}\; V  \; ,
\qquad
\lgl \;  a_0 \; \rgl_\ep \; = \; \frac{\ep}{\mu}\; \sqrt{V} \; e^{i\vp} \;  .
\ee
For uncondensed particles, one obtains the distribution
\be
\label{85}
\lgl \; a_k^\dgr \; a_k \; \rgl_\ep \; = \; \frac{1}{e^{\bt\om_k}-1}\; , \qquad
\om_k \; \equiv \; \frac{k^2}{2m} \; - \; \mu \; .
\ee

The number of condensed particles becomes
\be
\label{86}
 N_0 \; = \;  \lgl \; a_0^\dgr \; a_0 \; \rgl_\ep \; = \; 
\frac{z}{1-z} + \frac{\ep^2}{\mu^2} \; V \; .
\ee
Keeping in mind that the chemical potential is small allows us to accomplish the 
simplification
$$
\frac{z}{1-z} \; \simeq \; - \; \frac{T}{\mu} \qquad ( \mu \ra -0 ) \;  ,
$$
which, actually, is not principal. Therefore the number of condensed particles can be 
written as
\be
\label{87}
 N_0 \; = \; - \; \frac{T}{\mu}  +  \frac{\ep^2}{\mu^2}\; V \; .
\ee
Being interested in the fluctuations of condensed particles, we need to consider the
variance
\be
\label{88}
 {\rm var}(\hat N_0) \; = \; \lgl \; \hat N_0^2 \; \rgl -
\lgl \hat N_0 \; \rgl^2 \; = \; T \; \frac{\prt N_0}{\prt \mu} \; ,
\ee
which reads as 
\be
\label{89}
 {\rm var}(\hat N_0) \; = \;  \frac{T^2}{\mu^2} \; - \; 
\frac{2T\ep^2}{\rho \mu^3}\; N \;  .
\ee

In the method of quasi-averages, first, one has to take the thermodynamic limit for the 
relative quantity
\be
\label{90}
\lim_{N\ra\infty} \; \frac{{\rm var}(\hat N_0)}{N} \; = \; -\; 
\frac{2T}{\rho\mu^3} \; \ep^2 \; . 
\ee
After this, one has to sent $\varepsilon$ to zero, which gives
\be
\label{91}
\lim_{\ep\ra 0} \; \lim_{N\ra\infty} \; \frac{{\rm var}(\hat N_0)}{N} \; = \; 0 \;  .
\ee
Thus, not merely the spurious catastrophic condensate fluctuations do not exist, but the 
condensate fluctuations are absent at all.

In order to shed a bit more of light on the difference between calculations with gauge 
symmetry breaking and without it, let us notice the following. In the correct calculation
with the gauge symmetry breaking, the chemical potential, is defined by the equation 
$N = N_0 + N_1$, with the number of condensed particles (\ref{86}) and the number of 
uncondensed particles
\be
\label{92}
 N_1 \; = \; \sum_{k\neq 0} \frac{1}{e^{\bt\om_k}-1} \; = \; 
\frac{1}{\lbd_T^3} \; g_{3/2}(z) \; V \;  ,
\ee
where $\lambda_T \equiv \sqrt{2 \pi/m T}$ is the thermal wavelength and $g_n(z)$ is the
Bose-Einstein integral function \cite{Ziff_50}. From the equation
\be
\label{93}
\rho\; = \; - \; \frac{T}{\mu V} + \frac{\ep^2}{\mu^2} + 
\frac{1}{\lbd_T^3} \; g_{3/2}(z) \; ,
\ee
 using the expression
$$
 g_{3/2}(z) \; \simeq \; \zeta\left(\frac{3}{2}\right) + 
\Gm\left(-\; \frac{1}{2}\right) \; \sqrt{-\bt\; \mu} 
\qquad 
(\mu\ra -0) \;  ,
$$ 
where $\zeta$ is the Riemann zeta function,
we find
\be
\label{94}
\mu \; = \; -\; \frac{|\;\ep\;|}{\sqrt{\rho_0} } \; - \; \frac{T}{2\rho_0 V} \;  ,
\ee
where
\be
\label{95}
\rho_0 \; = \; \rho - \rho_1 \; , \qquad 
\rho_1 \; = \; \frac{1}{\lbd_T^3}\; g_{3/2}(1) \;   .
\ee
As is seen, the chemical potential, under the thermodynamic limit, is finite. Also, the 
main term, defining the number of condensed particles in equations (\ref{86}) and 
(\ref{93}) is the second term $\varepsilon^2/\mu^2$, while, if the symmetry is not 
broken, it does not appear at all and the number of condensed particles is given
by the first term $z/(1-z) = -T/\mu$.  

The conclusion on the absence of condensate fluctuations looks absolutely evident, 
and even trivial, if we remember that the method of quasi-averages is equivalent to 
the Bogolubov shift described in Sec. 5.6. The use of the Bogolubov shift (\ref{65}) 
tells us that the number-of-particle operator for condensed particles is just the 
unity operator factored with a constant,
\be
\label{96}
 \hat N_0 \; = \; N_0 \; \hat 1 \;  .
\ee
As is obvious, the variance for this operator is identically zero:
\be
\label{97}
 {\rm var}(\hat N_0) \; = \; 0 \;  .
\ee
Not merely there are no catastrophic condensate fluctuations, but the condensate 
fluctuations do not exist in principle. Their appearance is caused by incorrect
calculations in the frame of the grand canonical ensemble without gauge symmetry 
breaking.

\section{Ideal uniform Bose-condensed gas is unstable}

As is shown above, correct calculations, with gauge symmetry breaking, prove that 
there are no condensate fluctuations at all, so that all fluctuations are caused by 
uncondensed particles,
\be
\label{98}
 {\rm var}(\hat N) \; = \;  {\rm var}(\hat N_1) \;   .
\ee
For an ideal gas, the number of uncondensed particles is
\be
\label{99}
 N_1 \; = \; \frac{1}{\lbd_T^3} \; g_{3/2}(z) \; V \;  ,
\ee
where $g_n(z)$ is the ideal-gas Bose integral \cite{Ziff_50}. Particle fluctuations,
characterized by the variance
\be
\label{100}
 {\rm var}(\hat N_1) \; = \; T\; \frac{\prt N_1}{\prt\mu} \;   ,
\ee 
give 
\be
\label{101}
 {\rm var}(\hat N_1) \; = \; \frac{1}{\lbd_T^3} \; g_{1/2}(1) \; V \;  .
\ee

However the function $g_{1/2}(1)$ diverges, which hints on the system instability. In
order to find out how exactly the divergence occurs in a large but finite system, it 
is necessary either to study a discretized energy spectrum or to introduce a modified 
Bose function \cite{Yukalov_32} by taking into account that in a finite system there 
exists a minimal energy $\varepsilon_{min}$ that can be estimated as 
\be
\label{102}
 \ep_{min} \; = \; \frac{k^2_{min}}{2m} \qquad 
\left( k_{min}=\frac{2\pi}{L} \; , \; L = V^{1/3}\right) \;  .
\ee
The modified Bose integral is defined \cite{Yukalov_32} as
\be
\label{103}
 g_n(z) \; \equiv \; \frac{1}{\Gm(n)} \int_{u_0}^\infty \frac{z u^{n-1}}{e^u-z} \; du \;  ,
\ee
where
\be 
\label{104}
 u_0 \; \equiv \; \frac{\ep_{min}}{T} \; = \; \pi\; \frac{\lbd_T^2}{L^2}   
\ee
is a dimensionless minimal energy. Then one gets
$$
g_{1/2}(1) \; = \; \frac{2L}{\pi\lbd_T} \; = \; \frac{2}{\pi\lbd_T}\; V^{1/3} \;   .
$$

The variance (\ref{101}) becomes
\be
\label{105}
{\rm var}(\hat N_1) \; = \; \frac{m^2 T^2}{2\pi^3}\; V^{4/3} \; = \; 
\frac{2}{\pi\lbd_T^4}\; V^{4/3} \; ,   
\ee
which gives
\be
\label{106}
 \frac{{\rm var}(\hat N_1)}{N} \; = \;  \frac{m^2 T^2}{2\pi^3\rho}\; V^{1/3} \; = \; 
\frac{2}{\pi\rho\lbd_T^4}\; V^{1/3} \;  .
\ee
Thus the diverging compressibility 
\be
\label{107}
\varkappa_T \; = \; \frac{m^2T}{2\pi^3\rho^2}\; V^{1/3}
\ee
shows that the ideal uniform Bose-condensed gas is unstable. This is contrary to the 
normal not condensed Bose gas above the condensation point, where it is stable 
\cite{Yukalov_23}. The same expression (\ref{105}) is obtained by considering  the 
discrete quantum spectrum for a system in a box \cite{Ziff_50}.

The instability of the ideal Bose gas has also been noticed in its unphysical behavior 
depending on the boundary conditions, even for large systems. Thus it was found 
\cite{Vandevenne_56} that the ideal Bose gas is unstable with respect to boundary 
conditions, whose slight variation leads to a dramatic change of the spatial particle 
distribution, even in the thermodynamic limit. This is contrary to the behavior of 
realistic stable systems, for which the influence of boundary conditions disappears 
in the thermodynamic limit. Changing, for the ideal Bose gas, the boundary conditions 
can lead to the Bose-Einstein condensation from the bulk phenomenon to a strange 
surface effect, when the condensate is localized in a narrow domain in the vicinity 
of the system surface, being mainly concentrated at the corners of an infinite box. 

Fortunately, the instability of the ideal uniform Bose-condensed gas is not a great 
problem, since such a gas is just an unrealistic model, while any real gas does exhibit
some, at least minimal, interactions. The absolutely ideal gas and a gas with even 
infinitesimally weak interactions are principally different objects.

\section{Bose-Einstein condensation in interacting systems}

A system of interacting particles with Bose-Einstein condensate, for arbitrary interaction 
strength and temperature, is described by a self-consistent approach \cite{Yukalov_13,
Yukalov_16,Yukalov_21,Yukalov_23,Yukalov_57,Yukalov_58,Yukalov_59,Yukalov_60,Yukalov_61}.
This approach gives a gapless and conserving theory, resolving the Hohenberg--Martin
\cite{Hohenberg_63} dilemma stating that in a Bose system with the broken global gauge 
symmetry either there appears an unphysical gap in the spectrum or thermodynamic relations
become invalid. 

We start with the usual energy Hamiltonian with two-body interactions $\Phi({\bf r})$,
\be
\label{108}
 \hat H \; = \; 
\int \psi^\dgr(\br) \; \left( - \; \frac{\nabla^2}{2m}\right) \; \psi(\br) \; d\br 
+  \frac{1}{2} \int \psi^\dgr(\br) \; \psi^\dgr(\br') \; \Phi(\br-\br') \; 
\psi(\br') \; \psi(\br) \; d\br d\br' \; ,
\ee
in which $\psi({\bf r}) = \psi({\bf r},t)$ are field operators with Bose commutation 
relations. For short, the notation for time $t$ can be omitted, where this does not lead 
to confusion.

Aiming at describing a system with Bose condensate, one has to break the gauge symmetry.
The most convenient way for doing this is to resort to the Bogolubov shift 
\cite{Bogolubov_39,Bogolubov_40,Bogolubov_41}
\be
\label{109}
 \psi(\br,t) \; = \; \eta(\br,t) + \psi_1(\br,t) \;  ,
\ee
in which $\eta$ is the condensate wave function and $\psi_1$ is the operator of 
uncondensed particles. As is explained in Sec. 5.6, to avoid double counting, $\eta$ and
$\psi_1$ are mutually orthogonal, as in condition (\ref{69}). Recall that the Bogolubov 
shift is an exact canonical transformation, introduced for breaking the gauge symmetry, 
but not an approximation. 

The number of condensed particles is
\be
\label{110}
 N_0 \; = \; \lgl \; \hat N_0 \; \rgl \; , \qquad \hat N_0 \; = \; N_0 \; \hat 1 \; ,
\qquad
N_0 \; = \; \int |\; \eta(\br,t) \; |^2 \; d\br  
\ee
and the number of uncondensed particles is
\be
\label{111}
 N_1 \; = \; \lgl \; \hat N_1 \; \rgl \; , \qquad 
\hat N_1 \; = \; \int \psi_1^\dgr(\br) \; \psi_1(\br) \; d\br   .
\ee
In order that condition (\ref{68}) be valid, it is required that the Hamiltonian does not
contain the terms linear in $\psi_1$, which can be represented as the average
\be
\label{112}
\lgl \; \hat \Lbd \; \rgl \; = \; 0 \; , 
\qquad
\hat\Lbd \; = \; 
\int \left[\; \lbd(\br) \; \psi_1^\dgr(\br) + \lbd^*(\br) \; \psi_1(\br) \; \right] \; d\br \; .
\ee
The above conditions are satisfied with the grand Hamiltonian 
\be
\label{113}
H \; = \; \hat H - \mu_0 N_0 - \mu_1 \hat N_1 - \hat \Lbd \;   ,
\ee
with the Lagrange multipliers $\mu_0$, $\mu_1$, and $\lambda({\bf r})$. 
 
Since the condensate wave function and the operator of uncondensed particles are 
mutually independent orthogonal functional variables, one has to write two equations
of motion, for the condensate function,
\be
\label{114}
 i\; \frac{\prt}{\prt t}\; \eta(\br,t) \; = \; 
\left\lgl\; \frac{\dlt H}{\dlt\eta^*(\br,t)} \; \right\rgl \; ,
\ee
and for the operator of uncondensed particles,
\be
\label{115}
 i\; \frac{\prt}{\prt t}\; \psi_1(\br,t) \; = \; 
\frac{\dlt H}{\dlt\psi_1^\dgr(\br,t)} \; .
\ee
It is useful to keep in mind the equality \cite{Yukalov_16}
\be
\label{116}
\frac{\dlt H}{\dlt\psi_1^\dgr(\br,t)} \; = \; [\; \psi_1(\br,t)\; , \; H\; ] \; .
\ee
The system chemical potential reads as
\be
\label{117}
\mu \; = \; \mu_0 n_0 + \mu_1 n_1 \; ,
\ee
where the concentrations of condensed and uncondensed particles are
\be
\label{118}
n_0 \; = \; \frac{N_0}{N} \; , \qquad n_1 \; = \; \frac{N_1}{N} \; .
\ee

In what follows, we need the first-order reduced density matrix
\be
\label{119}
\rho_1(\br,\br') \; \equiv \; \lgl \; \psi_1^\dgr(\br') \; \psi_1(\br) \; \rgl
\ee
and the anomalous average
\be
\label{120}
 \sgm_1(\br,\br') \; = \; \lgl \; \psi_1(\br') \; \psi_1(\br) \; \rgl \; .
\ee

The density of particles 
\be
\label{121}
\rho(\br) \; = \; \rho_0(\br) + \rho_1(\br)
\ee
consists of the condensate density 
\be
\label{122}
 \rho_0(\br) \; = \; |\; \eta(\br)\; |^2
\ee
and the density of uncondensed particles  
\be
\label{123}
 \rho_1(\br) \; = \; \rho_1(\br,\br) \; = \; 
\lgl \; \psi_1^\dgr(\br) \; \psi_1(\br) \; \rgl \; .
\ee
The anomalous average (\ref{120}) defines the density
\be
\label{124}
 \rho_\pi(\br,\br') \; = \; |\; \sgm_1(\br,\br')\; |^2
\ee
of pair-correlated particles, one of which being at the point ${\bf r}$ and the other, 
at ${\bf r'}$. 

If one employs in the right-hand side of equation (\ref{114}) the averaging over the 
vacuum state, that is, over the coherent state satisfying the condition
\be
\label{125}
 \psi_1(\br) \; |\; \eta\; \rgl \; = \; 0 \; , \qquad 
\psi(\br) \; |\; \eta\; \rgl \; = \; \eta(\br) \; |\; \eta\; \rgl \; ,
\ee
then one obtains the coherent approximation for the condensate wave function equation, 
\be
\label{126}
 i \; \frac{\prt}{\prt t} \; \eta(\br,t) \; = \; 
\left( - \; \frac{\nabla^2}{2m} \; - \; \mu_0 \right) \; \eta(\br,t) + 
\int \Phi(\br-\br') \; |\; \eta(\br',t)\; |^2 \; d\br' \; \eta(\br,t) .
\ee
This equation was advanced by Bogolubov \cite{Bogolubov_64} in 1949 and has been 
republished numerous times (see, e.g., \cite{Bogolubov_39,Bogolubov_40,Bogolubov_41}). 
By its form, equation (\ref{126}) is a nonlinear Schr\"{o}dinger equation \cite{Malomed_65}.

In the more general Hartree-Fock-Bogolubov (HFB) approximation, the evolution equation 
(\ref{114}) results in the equation for the condensate wave function that reads as
$$
i \; \frac{\prt}{\prt t} \; \eta(\br,t) \; = \; 
\left( - \; \frac{\nabla^2}{2m} \; - \; \mu_0 \right) \; \eta(\br,t) \; + 
$$
\be
\label{127}
 +\;  
\int \Phi(\br-\br') \; [\; \rho(\br') \; \eta(\br) + \rho_1(\br,\br') \; \eta(\br') +
\sgm_1(\br,\br')\; \eta(\br')\; ] \; d\br' \; .
\ee
As is evident, the coherent approximation (\ref{126}) for the condensate-function 
equation follows from the HFB approximation (\ref{127}) when neglecting the uncondensed
particles by sending $\rho_1$ and $\sigma_1$ to zero. 

For a uniform system, passing to the momentum representation for the field operator,
$$
 \psi_1(\br) \; = \; \frac{1}{\sqrt{V} } \sum_{k\neq 0} a_k \; e^{i\bk \cdot\br} \;  ,
$$ 
and for the interaction potential,
$$
 \Phi(\br) \; = \; \frac{1}{V} \sum_k \Phi_k \; e^{i\bk \cdot\br} \; , 
\qquad
\Phi_k \; = \; \int  \Phi(\br) \; e^{- i\bk \cdot\br} \; d\br \;  ,
$$
we come to the constant densities
\be
\label{128}
\rho_0(\br) \; = \; \rho_0 \; , \qquad \rho_1(\br) \; = \; \rho_1 \; , \qquad 
\rho(\br) \; = \; \rho \; = \; \rho_0 + \rho_1 \;  .   
\ee
The normal,
\be
\label{129}
 \rho_1(\br,\br') \; = \; 
\frac{1}{V} \sum_{k\neq 0} n_k \; e^{i\bk\cdot(\br-\br')} \; ,
\ee
and anomalous,
\be
\label{130}
\sgm_1(\br,\br') \; = \; 
\frac{1}{V} \sum_{k\neq 0} \sgm_k \; e^{i\bk\cdot(\br-\br')} \;  ,
\ee 
averages depend on the difference ${\bf r} - {\bf r'}$. 

In the HFB approximation, the momentum representation for the normal and anomalous 
averages acquire the form
\be
\label{131}
 n_k \; = \; \lgl \; a_k^\dgr \; a_k \; \rgl \; = \; 
\frac{\om_k}{2\ep_k} \; \coth\left( \frac{\ep_k}{2T}\right) - \; \frac{1}{2}   
\ee
and, respectively,
\be
\label{132}
\sgm_k \; = \; \lgl \; a_k \; a_{-k} \; \rgl \; = \; 
- \; \frac{\Dlt_k}{2\ep_k} \; \coth\left( \frac{\ep_k}{2T}\right) \;  .
\ee
The spectrum of collective excitations reads as
\be
\label{133}
 \ep_k \; = \; \sqrt{\om_k^2 -\Dlt_k^2} \;  ,
\ee
where
\be
\label{134}
\om_k \; = \; \frac{k^2}{2m} + \Dlt + \rho_0 \; ( \Phi_k - \Phi_0 ) +
\frac{1}{V} \sum_{p\neq 0} n_p \; (\Phi_{k+p} - \Phi_p ) 
\ee
and 
\be
\label{135}
 \Dlt_k \; = \; \rho_0 \; \Phi_k +  \frac{1}{V} \sum_{p\neq 0} \sgm_p \; \Phi_{k+p} \; ,
\ee
while
\be
\label{136}
\Dlt \; \equiv \; \lim_{k\ra 0} \Dlt_k \; = \; \rho_0 \; \Phi_0 + 
\frac{1}{V} \sum_{p\neq 0} \sgm_p \; \Phi_p \;   .
\ee
The chemical potentials are
\be
\label{137}
\mu_0 \; = \; 
\rho \; \Phi_0 +  \frac{1}{V} \sum_{k\neq 0} (n_k + \sgm_k) \; \Phi_k \; ,
\qquad
\mu_1 \; = \; 
\rho \; \Phi_0 +  \frac{1}{V} \sum_{k\neq 0} (n_k - \sgm_k) \; \Phi_k \;  .
\ee

In the long-wave limit, the spectrum (\ref{133}) is acoustic,
\be
\label{138}
 \ep_k \; \simeq \; ck \qquad ( k \ra 0 ) \;  ,
\ee
where the sound velocity is
\be
\label{139}
c \; = \; \sqrt{\frac{\Dlt}{m^*} } \;   ,
\ee
with the effective mass 
\be
\label{140}
 m^* \; = \; \frac{m}{1+ \frac{m}{V}\sum_p(n_p-\sgm_p)\Phi_p''}
\qquad
\left( \Phi_p'' \equiv \frac{\prt^2\Phi_p}{\prt p^2} \right) \;  .
\ee
The spectrum is gapless, as it should be according to the Hugenholtz-Pines theorem
\cite{Hugenholtz_66}.

\section{Particle fluctuations in interacting systems}

The number-of-particle variance (\ref{27}) is expressed through the pair correlation 
function that for a uniform system reads as 
\be
\label{141}
g(\br-\br') \; = \; \frac{1}{\rho^2} \; \rho_2(\br,\br',\br,\br') \;  .
\ee
In the HFB approximation, one has
\be
\label{142}
 g(\br_1-\br_2) \; = \;  1 + \frac{2\rho_0}{\rho^2} \; [\; \rho_1(\br_1,\br_2) +
\sgm_1(\br_1,\br_2) \; ] + \frac{1}{\rho^2} 
\left[ \; |\; \rho_1(\br_1,\br_2)\; |^2 + |\; \sgm_1(\br_1,\br_2)\; |^2 \; \right] \; .
\ee
This gives
\be
\label{143}
 \int [\; g(\br) -1 \; ] \; d\br \; = \;  \frac{2\rho_0}{\rho^2} \;
\lim_{k\ra 0} ( n_k + \sgm_k) +
\frac{1}{\rho^2} \int ( n_k^2 + \sgm_k^2 ) \; \frac{d\bk}{(2\pi)^3}  \; .
\ee

The long-wave limit of expressions (\ref{131}) and (\ref{132}), keeping in mind the
spectrum (\ref{138}), results in the momentum distribution
\be
\label{144}
n_k \; \simeq \; \frac{T\Dlt_k}{\ep_k^2} + \frac{\Dlt_k}{12T} +  \frac{T}{2\Dlt_k} \; - \;
\frac{1}{2} + \left(  \frac{\Dlt_k}{3T} \; - \;  \frac{T}{\Dlt_k} \; - \;
\frac{\Dlt_k^3}{90T^3} \right) \; \frac{\ep_k^2}{8\Dlt_k^2} \; ,
\ee
and the anomalous average 
\be
\label{145}
\sgm_k \; \simeq \; -\; \frac{T\Dlt_k}{\ep_k^2} \; - \;  \frac{\Dlt_k}{12T} +  
\frac{\Dlt_k\ep_k^2}{720 T^3} \qquad ( \ep_k \ra 0 ) \; .
\ee
Then the first term in (\ref{143}) yields
\be
\label{146}
\lim_{k\ra 0} (n_k + \sgm_k) \; = \; 
\frac{1}{2} \; \left( \frac{T}{\Dlt} \; - \; 1 \right) \;   .
\ee
 
Yet, the second term in (\ref{143}) is divergent. To show the character of the 
divergence, it is necessary to specify the lower limit of the integration by taking 
account of the system finiteness, which requires to set the minimal lower momentum 
$k_{min} = 2 \pi/L = 2 \pi /V^{1/3}$. Then
\be
\label{147}
 \frac{1}{\rho} \int (n_k^2 + \sgm_k^2) \; \frac{d\bk}{(2\pi)^3} \; = \; 
\frac{2}{\pi\rho\lbd_T^4}\; V^{1/3} \;  ,
\ee
where $\lambda_T \equiv \sqrt{2 \pi/ m^* T}$. The same divergence occurs if one makes
calculations for a quantum system with discrete spectrum.
    
As is seen, this divergence, if it would be not a spurious artefact, would mean that 
the particle fluctuations are thermodynamically anomalous, since the resulting 
compressibility would also be divergent. However this divergence is not physical, but 
it is caused by the mistake of going beyond the approximation applicability. Really, 
the HFB approximation reduces the initial Hamiltonian to the form that is quadratic 
with respect to the operators of uncondensed particles. In that sense, this is the 
second-order approximation. At the same time, for describing particle fluctuations, 
one needs to consider the terms of fourth order with respect to the operators of 
uncondensed particles. Straightforward mathematics tells us that calculating the 
fourth-order expressions using a second-order approximation is not self-consistent 
and can lead to abnormal results including divergences. This concerns the use of the 
HFB approximation for treating the fourth-order terms in the number-of-particle variance, 
as well as the use of the Bogolubov approximation or equivalent to it the hydrodynamic 
approximation that are also of second order. To be self-consistent, one has to 
disregard in the final result the terms of fourth order with respect to the operators 
of uncondensed particles, that is the terms containing such expressions as $n_k^2$, 
$\sigma_k^2$, and $n_k \sigma_p$. Then we obtain the variance
\be
\label{148}
{\rm var}(\hat N) \; = \; \frac{T}{m^* c^2} \; N \;   ,
\ee
hence the compressibility
\be
\label{149}
\varkappa_T \; = \; \frac{1}{\rho m^* c^2} \;   .
\ee
As is evident, no anomalous nonthermodynamic fluctuations arise when one accomplishes
correct self-consistent calculations. More discussions and explanations are given in
Refs. \cite{Yukalov_9,Yukalov_16,Yukalov_17,Yukalov_34,Yukalov_35,Yukalov_57,Yukalov_67,
Yukalov_68,Yukalov_69}.

It is possible to note that the divergence in expression (\ref{147}) is the same as in
formula (\ref{106}) for the ideal gas. The main difference is that for the ideal gas no
approximation has been necessary, and the divergence of the compressibility (\ref{107})
is real and does signify the ideal gas instability. While for the interacting particles, 
the divergence (\ref{147}) is unphysical, being due to incorrect not self-consistent 
calculations caused by going beyond the approximation applicability.  

In addition, it looks to be absolutely clear that the divergence of the compressibility 
cannot happen for really existing equilibrium systems, since then, according to relations 
(\ref{13}) and (\ref{28}), this would imply that the sound velocity is zero and the 
structural factor is infinite.    

Technically, the divergences in the terms above the second order with respect to the 
products of the field operators of uncondensed particles are to be suppressed by
taking account of the attenuation effects appearing in the orders higher than second.
As is known \cite{Beliaev_70,Beliaev_71,Giorgini_72} the spectrum of collective 
excitations in a Bose system acquires complex-valued sifts describing damping rates, 
as compared to the lower-order approximations, where the spectrum is purely real.  

It is possible to note that imposing on the system a random external potential leads
to the increase of particle fluctuations \cite{Yukalov_98} due to the suppressed
sound velocity, but no nonthermodynamic divergence arises.

\section{Ideal gas in rectangular box}

Bose systems in a rectangular box are not merely an interesting theoretical object, 
but they can also be realized experimentally inside box-shaped traps confining 
interacting Bose gases \cite{Gaunt_73,Navon_74,Lopes_75}. In Sec. 8, it is shown 
that the ideal uniform Bose-condensed gas in three dimensions is unstable. This 
analysis can be generalized to the study of stability of this gas in arbitrary 
spatial dimensions \cite{Yukalov_16,Yukalov_17,Yukalov_32,Yukalov_76,Yukalov_77}.   
 
For a finite $d$-dimensional system, it is possible to introduce the characteristic 
length
\be
\label{150}
 L \; = \; V^{1/d} \; = \; \left( \frac{N}{\rho}\right)^{1/d} \;  ,
\ee
which defines the minimal wave vector $k_{min}$ and the minimal energy,
\be
\label{151}
\ep_{min} \; = \; \frac{k_{min}^2}{2m} \; =\; \frac{2\pi^2}{mL^2} \qquad
\left( \; k_{min} = \frac{2\pi}{L} \; \right) \;   .
\ee
The dimensionless minimal energy 
\be
\label{152}
 u_0 \; \equiv \; \frac{\ep_{min}}{T} \; =\; \pi\; \frac{\lbd_T^2}{L^2} \qquad
\left(\; \lbd_T = \sqrt{\frac{2\pi}{mT} } \;\right) 
\ee
is assumed to be small, as far as $L \ra \infty$.

For a large, although finite, system, with $N \gg 1$, it is admissible to employ the 
integral representation by introducing the modified Bose integral \cite{Yukalov_32}
\be
\label{153}
g_n(z) \; \equiv \; 
\frac{1}{\Gm(n)} \int_{u_0}^\infty \frac{z u^{n-1}}{e^u-z} \; du \; .
\ee
The total number of particles is given by the expressions
\be
\label{154}
  N \; = \; N_0 + N_1 \; , \qquad N_1 \; = \; \frac{V}{\lbd_T^d}\; g_{d/2}(z) \; .
\ee
Recall that $g_n(1) = \zeta(n)$ for $n > 1$. Everywhere below, we use the notation 
$g_n(z)$ keeping in mind the modified Bose integral (\ref{153}).

\subsection{Condensation temperature in a box}

At the temperature of Bose-Einstein condensation, the chemical potential becomes zero,
hence the fugacity $z = 1$ and the condensation temperature is defined by the condition
\be
\label{155}
 N \; = \; \frac{V}{\lbd_T^d}\; g_{d/2}(1) \qquad ( T = T_c ) \;  ,
\ee
which results in the temperature
\be
\label{156}
T_c \; =\; \frac{2\pi}{m} \; \left[\; \frac{\rho}{g_{d/2}(1)} \; \right]^{2/d} \;   .
\ee
By introducing the characteristic kinetic energy
\be
\label{157}
 E_d \; \equiv \; \frac{\rho^{2/d}}{2m} \;  ,
\ee
the condensation temperature can be written as
\be
\label{158}
 T_c \; =\; 4\pi E_d \; \left[\; \frac{1}{g_{d/2}(1)}\; \right]^{2/d} \;  .
\ee 
In particular, in one dimension
\be
\label{159}
 T_c \; = \; \pi\; \sqrt{\ep_{min} \; E_1} \qquad ( d = 1) \;  ,
\ee
in two dimensions
\be
\label{160}
 T_c \; = \; \frac{4\pi E_2}{\ln(4\pi E_2/\ep_{min})}  \qquad ( d = 2) \;  ,
\ee
and in three dimensions
\be
\label{161}
  T_c \; = \; 4\pi E_3 \; \left[\; \frac{1}{\zeta(3/2)} \; \right]^{2/3}  \qquad 
( d = 3) \;   .
\ee
 
Due to the form (\ref{151}) of the minimal energy, there is the following scaling of
the critical temperature for large $N \gg 1$:
$$
T_c \;  \propto \; \frac{1}{N} \qquad ( d = 1) \; ,
$$
$$
T_c \;  \propto \; \frac{1}{\ln N} \qquad ( d = 2) \; ,
$$
\be
\label{162}
T_c \;  \propto \; const \qquad ( d = 3 ) \; .
\ee
The condensation temperature for $d \leq 2$ tends to zero, as $N \ra \infty$, but 
remains finite for $d \geq 3$.

\subsection{Fluctuations in normal gas $(T > T_c)$}

At the temperature above the critical point, there exist only normal particles. The 
chemical potential is negative, hence $z < 1$. The relative number-of-particle 
variance is
\be
\label{163}
 \frac{{\rm var}(\hat N)}{N} \; = \; 
\frac{z}{\rho\lbd_T^d} \; \frac{\prt}{\prt z} \; g_{d/2}(z) \;  .
\ee
Using the inequality $u_0 \ll 1$, we find for one dimension
\be
\label{164}
\frac{{\rm var}(\hat N)}{N} \; = \; -\; \frac{2z}{(1-z)^2 N} 
\qquad 
( d = 1) \; ,
\ee
for two dimensions
\be
\label{165}
\frac{{\rm var}(\hat N)}{N} \; = \; -\; \frac{\pi z}{(1-z)^2 N} 
\qquad 
( d = 2)  \; ,
\ee
and for three dimensions
\be
\label{166}
\frac{{\rm var}(\hat N)}{N} \; = \; \frac{1}{\rho \lbd_T^3} \; g_{1/2}(z) 
\qquad 
( d = 3) \; .
\ee
The negative values of the relative variance for the low-dimensional systems, with 
$d = 1$ and $d = 2$, imply that such systems are unstable. The gas is stable in three 
dimensions and in all higher dimensions $d > 2$. Thus the ideal normal gas is 
stable for
\be
\label{167}
d \; > \; 2 \qquad ( T > T_c ) \;  .
\ee

\subsection{Fluctuations in condensed gas $(T < T_c)$}

In a system with Bose-Einstein condensate, the total number of particles is the sum 
(\ref{154}). As is explained in detail in the above sections, in the correct description 
of the condensed systems, where the global gauge symmetry is broken, there are no 
fluctuations of the condensate, so that only uncondensed particles contribute to the
system particle fluctuations, ${\rm var}(\hat N) = {\rm var}(\hat N_1)$. The relative 
variance can be found from the formula
\be
\label{168}
 \frac{{\rm var}(\hat N)}{N} \; = \;
 \frac{1}{\rho\lbd_T^d} \; \lim_{z\ra 1} \; \frac{\prt}{\prt z} \; g_{d/2}(z) \; = \;
\frac{1}{\rho\lbd_T^d} \; \left[\; g_{(d-2)/2}(1) + 
\frac{u_0^{(d-4)/2}}{\Gm(d/2)} \; \right] \; .
\ee
This yields the following expressions for one dimension,
\be
\label{169}
 \frac{{\rm var}(\hat N)}{N} \; = \; \frac{2}{3\sqrt{\pi}\; \rho \lbd_T} \; u_0^{-3/2} \; = \; 
\frac{2L^3}{3\pi^2 \rho \lbd_T^4} 
\qquad 
( d = 1) \;  ,
\ee
for two dimensions,
\be
\label{170}
\frac{{\rm var}(\hat N)}{N} \; = \; \frac{1}{\rho \lbd_T^2} \; u_0^{-1} \; = \; 
\frac{L^2}{\pi \rho \lbd_T^4} 
\qquad 
( d = 2) \; ,
\ee
for three dimensions,
\be
\label{171}
\frac{{\rm var}(\hat N)}{N} \; = \; \frac{4}{\sqrt{\pi} \rho \lbd_T^3} \; u_0^{-1/2} \; = \; 
\frac{4L}{\pi \rho \lbd_T^4} 
\qquad 
( d = 3) \;  , 
\ee
for four dimensions
\be
\label{172}
\frac{{\rm var}(\hat N)}{N} \; = \; \frac{1}{\rho \lbd_T^4} \;\ln  u_0 \; = \; 
\frac{1}{\rho \lbd_T^4} \; \ln \; \left( \pi \; \frac{\lbd_T^2}{L^2} \right)
\qquad 
( d = 4) \;  ,
\ee
and for five dimensions
\be
\label{173}
 \frac{{\rm var}(\hat N)}{N} \; = \; \frac{1}{\rho \lbd_T^5} \; g_{3/2}(1)
\qquad
( d = 5) \;   .
\ee

Taking account of the value $u_0$ in (\ref{152}), one has the scaling at large $N$ as
$$
\frac{{\rm var}(\hat N)}{N} \; \propto \; N^3 \qquad ( d = 1) \; ,
$$
$$
\frac{{\rm var}(\hat N)}{N} \; \propto \; N \qquad ( d = 2) \; ,
$$
$$
\frac{{\rm var}(\hat N)}{N} \; \propto \; N^{1/3} \qquad ( d = 3) \; ,
$$
$$
\frac{{\rm var}(\hat N)}{N} \; \propto \; \ln N \qquad ( d = 4) \; ,
$$
\be
\label{174}
 \frac{{\rm var}(\hat N)}{N} \; \propto \; const \qquad ( d = 5) \;  .
\ee

The system is stable for the dimensions
\be
\label{175}
d \; > \; 4 \qquad ( T < T_c ) \; ,
\ee
where the condensate fraction $n_0 \equiv N_0/N$ behaves as
$$
 n_0 \; = \; 1 - \left( \frac{T}{T_c}\right)^{d/2} \qquad ( T \leq T_c ) \;  .
$$

\section{Ideal gas in a power-law potential}

The power-law potential representing a trap can be written as
\be
\label{176}
U(\br) \; = \; 
\sum_{\al=1}^d \frac{\om_\al}{2} \; \left| \; \frac{r_\al}{l_\al} \; \right|^{n_\al} 
\qquad
\left( l_\al \equiv \frac{1}{\sqrt{m\om_\al} } \right) \;  .
\ee
The trap can be characterized by the effective frequency 
\be
\label{177}
  \om_0 \; \equiv \; \left( \prod_{\al=1}^d \om_\al\right)^{1/d} \; = \; \frac{1}{m l_0^2}  
\ee
and effective length
\be
\label{178}
 l_0 \; \equiv \; \left( \prod_{\al=1}^d l_\al\right)^{1/d} \; = \; \frac{1}{\sqrt{m \om_0} }   .
\ee

Setting the minimal energy as
\be
\label{179}
 \ep_{min} \; = \; \frac{1}{2} \; \om_0 \;  ,
\ee
the dimensionless minimal energy is taken in the form
\be
\label{180}
u_0 \; \equiv \; \frac{\ep_{min}}{T} \; = \; \frac{\om_0}{2T} \; .
\ee

A characteristic quantity appearing in the study of power-law potentials is the confining
dimension \cite{Yukalov_32}
\be
\label{181}
 D \; \equiv \; \frac{d}{2} + \sum_{\al=1}^d \frac{1}{n_\al} \;  .
\ee
The number of uncondensed particles, under the semiclassical approximation, reads as
\be
\label{182}
N_1 \; = \; \frac{T^D}{\gm_D} \; g_D(z) \;   ,
\ee
where
\be
\label{183}
\gm_D \; \equiv \; \frac{\pi^{d/2}}{2^D} \prod_{\al=1}^d 
\frac{\om_\al^{1/2+1/n_\al} }{\Gm(1+1/n_\al)} \;   .
\ee
The function $g_n(z)$ is the modified Bose integral that is the same as in (\ref{153}), 
except that the lower limit in the integral is set as $u_0$ defined in (\ref{180}).

\subsection{Condensation temperature in a power-law potential}

At the condensation temperature, one has $\mu = 0$, hence $z =1$, and $N_1 = N$. Then 
(\ref{182}) yields the temperature
\be
\label{184}
T_c \; = \; \left[ \; \frac{\gm_D N}{g_D(1)} \; \right]^{1/D} \;  .
\ee

For $D < 1$, the spatial dimension can be nothing except one, $d = 1$. Then the critical
temperature is
\be
\label{185}
T_c \; = \; \frac{\sqrt{\pi} (1-D)\Gm(D)}{\Gm(1+1/n)} \; \left( \frac{\om_0}{2}\right) \; N
\qquad ( D < 1 , ~ d = 1) \; .
\ee
For the confining dimension $D = 1$, one has $d = 1$ and $n = 2$. Then for the critical
temperature it follows
\be
\label{186}
T_c \; = \; \frac{N\om_0}{\ln(2T_c/\om_0)} 
\qquad ( D = d = 1, ~ n = 2) \; .
\ee
For large $N$, one has 
$$
 \ln \left( \frac{2T_c}{\om_0} \right) \; \ll \; 2 N \;  ,
$$ 
so that (\ref{186}) can be transformed to the expression
\be
\label{187}
T_c \; = \; \frac{N\om_0}{\ln( 2N)} 
\qquad ( D = d = 1, ~ n = 2) \;   .
\ee
It should be stressed that exactly the same critical temperature (\ref{186}) is obtained
\cite{Ketterle_78} by purely quantum-mechanical calculations with a discrete spectrum. 

When $D > 1$, the critical temperature is
\be
\label{188}
T_c \; = \; \left[ \; \frac{\gm_D N}{\zeta(D)} \; \right]^{1/D}  \qquad ( D > 1) \;   .
\ee
In particular, for harmonic traps, when 
$$
\gm_D \; = \; \om_0^d \qquad ( D = d , ~ n_\al = 2) \;   ,
$$
the temperature becomes
\be
\label{189}
 T_c \; = \; \om_0 \; \left[ \; \frac{N}{\zeta(d)} \; \right]^{1/d}  \qquad 
( D = d > 1, ~ n_\al = 2) \;  .
\ee

For large $N$, there exists the thermodynamic limit (\ref{2}), where as an extensive 
observable quantity one can take the average energy $E_N$, for which we have
\be
\label{190}
E_N \; = \; \frac{D}{\gm_D} \; g_{D+1}(z) \; T^{D+1} \;   .
\ee
Therefore the thermodynamic limit
\be
\label{191}
N \; \ra \; \infty \; , \qquad E_N \; \ra \; \infty \; , \qquad 
\frac{E_N}{N} \; \ra \; const
\ee
acquires the form
\be
\label{192}
N \; \ra \; \infty \; , \qquad \gm_D \; \ra \; 0 \; , \qquad 
N \; \gm_D \; \ra \; const \; .
\ee
For equipower potentials, where $n_\alpha = n$ and
$$
 \gm_D \; = \; \frac{\pi^{d/2}}{\Gm^d(1+1/n)} \; \left( \frac{\om_0}{2} \right)^D 
\qquad  ( n_\al = n) \; ,
$$
this translates to
\be
\label{193}
 N \; \ra \; \infty \; , \qquad \om_0 \; \ra \; 0 \; , \qquad 
N  \om_0^D \; \ra \; const  .
\ee

In the thermodynamic limit, the condensation temperature scales as
$$
T_c \; \propto \; \frac{1}{N^{1/D-1}} \qquad ( D < 1 ) \; 
$$
$$
T_c \; \propto \; \frac{1}{\ln N} \qquad ( D = 1 ) \; 
$$
\be
\label{194}
T_c \; \propto \; const \qquad ( D > 1 ) \; .
\ee

\subsection{Fluctuations above condensation temperature $(T > T_c)$}

In the region of temperatures above the condensation point particle fluctuations
are described by the relative variance
\be
\label{195}
 \frac{{\rm var}(\hat N)}{N} \; = \; 
\frac{T^D}{N\gm_D} \; z \; \frac{\prt}{\prt z} \; g_D(z) \;  .
\ee
Depending on the confining dimension, we find for $D < 1$ 
\be
\label{196} 
 \frac{{\rm var}(\hat N)}{N} \; = \; -\; \frac{z T^D}{(1-z)^2\Gm(1+D)N\gm_D} \; 
\left( \frac{\om_0}{2T} \right)^D  \qquad ( D < 1 ) \;  ,
\ee
for $D = 1$
\be
\label{197}
  \frac{{\rm var}(\hat N)}{N} \; = \; -\; \frac{z T}{(1-z)^2 N\gm_1} \; 
\left( \frac{\om_0}{2T} \right) \qquad ( D = 1) \;  ,
\ee
and for $D > 1$
\be
\label{198}
  \frac{{\rm var}(\hat N)}{N} \; = \; \frac{T^D}{N\gm_D} \; 
g_{D-1}(z) \qquad ( D > 1) \;   .
\ee

As we see, for $D \leq 1$ the variance is negative, hence the compressibility is negative,
which implies the system instability. The condition of stability for the confining 
dimension is
\be
\label{199}
  D \; > \; 1 \qquad ( T > T_c) \; .
\ee
For the spatial dimension $d$ this reads as
\be
\label{200}
 \frac{d}{2} + \sum_{\al=1}^d \frac{1}{n_\al} \; > \; 1 \qquad (T > T_c) \;  .
\ee

Sending $n_\alpha$ to infinity returns us back to the rectangular box, where the 
stability condition above $T_c$ is (\ref{167}) that agrees with condition (\ref{200}).

\subsection{Fluctuations below condensation temperature $(T < T_c)$}

Below the temperature of condensation, $\mu = 0$, $z = 1$, and $N = N_0 + N_1$. 
Following the correct way of describing the condensate, one has to break the global 
gauge symmetry, which leads to zero condensate fluctuations, so that all particle 
fluctuations in the system are due to uncondensed particles, 
${\rm var}(\hat N) = {\rm var}(\hat N_1)$. This gives the relative variance
\be
\label{201}
 \frac{{\rm var}(\hat N)}{N} \; = \; \frac{T^D}{N\gm_D} \; \lim_{z\ra 1} \;
\frac{\prt}{\prt z} \; g_D(z) \;  .
\ee

From here, one obtains for $D < 2$
\be
\label{202}
 \frac{{\rm var}(\hat N)}{N} \; = \; \frac{T^D}{N\gm_D} \; \left[
\frac{1}{(2-D)\Gm(D-1)} + \frac{1}{\Gm(D)} \; \right] \;
\left( \frac{2T}{\om_0} \right)^{2-D} \qquad ( D < 2) \;  ,
\ee
for $D = 2$,
\be
\label{203}
 \frac{{\rm var}(\hat N)}{N} \; = \; 
\frac{T^2}{N\gm_2} \; \ln \left( \frac{2T}{\om_0} \right) 
\qquad
( D = 2) \;  ,
\ee
and for $D > 2$, we get
\be
\label{204}
 \frac{{\rm var}(\hat N)}{N} \; = \; 
\frac{T^D}{N\gm_G} \; \zeta(D-1)
\qquad
( D > 2) \;    .
\ee

The scaling of the variance with respect to the number of particles gives
$$
\frac{{\rm var}(\hat N)}{N} \; \propto \; N^{(2-D)/D} \qquad ( D < 2 ) \; ,
$$
$$
\frac{{\rm var}(\hat N)}{N} \; \propto \; \ln N \qquad ( D = 2 ) \; ,
$$
\be
\label{205}
 \frac{{\rm var}(\hat N)}{N} \; \propto \; const \qquad ( D = 3 ) \;  .
\ee
The condition of stability becomes
\be
\label{206}
D \; > \; 2 \qquad ( T < T_c ) \;   ,
\ee
or explicitly
\be
\label{207}
 \frac{d}{2} + \sum_{\al=1}^d \frac{1}{n_\al} > 2 \qquad ( T < T_c) \;  .
\ee

In the case of harmonic traps, when $n_\alpha = 2$, so that $D = d$ and 
$\gamma_d = \omega^d$, the stability conditions (\ref{206}) and (\ref{207}) reduce 
to $d > 2$. Then the relative variance is
\be
\label{208}
 \frac{{\rm var}(\hat N)}{N} \; = \; 
\frac{\zeta(d-1)}{\zeta(d)} \; \left( \frac{T}{T_c} \right)^d 
\qquad 
( n_\al = 2 , ~ d > 2 )\;   .
\ee
This means that in a harmonic trap the ideal condensed Bose system is stable starting 
from $d = 3$. The ideal gas in one- and two-dimensional harmonic traps is unstable.
In experiments, one often considers quasi-one-dimensional and quasi-two-dimensional 
traps that, strictly speaking, are actually three-dimensional, but tightly confined 
in one or two directions. In addition, the gases studied in experiments are always 
interacting, at least weakly. When the stability condition (\ref{206}) is valid, 
the ideal Bose gas exhibits the condensate fraction
$$
 n_0 \; = \; 1 - \left( \frac{T}{T_c} \right)^D \qquad ( T \leq T_c) \;  .
$$

\section{Fluctuations of composite observables}

Often it is required to consider the variance of a composite observable that is a sum 
of other observables. The following theorem connects the behavior of the composite 
observable with that of the partial terms of the sum \cite{Yukalov_57,Yukalov_68}. 
For generality, under observables we keep in mind quantum operators, although the same 
results are valid for classical random variables.  

\vskip 2mm

{\bf Theorem}. Let us consider a composite operator
\be
\label{209}
\hat A \; = \; \sum_i \hat A_i
\ee
being a sum of linearly independent terms. And let, under the parameter $N \gg 1$, the 
related variances behave as 
\be
\label{210}
{\rm var}(\hat A) \; = \; c N^\al \; , 
\qquad  
{\rm var}(\hat A_i) \; = \; c_i N^{\al_i} \;  ,
\ee
with real-valued powers $\alpha_i$ and coefficients $c_i$ not depending on $N$. Then 
\be
\label{211}
 \al \; = \; \sup_i \al_i \;  .
\ee

\vskip 2mm 

{\bf Proof}. The variance of the composite operator can be written as
\be
\label{212}
{\rm var}(\hat A) \; = \; \sum_i {\rm var}(\hat A_i) + 
\sum_{i\neq j} {\rm cov}(\hat A_i, \; \hat A_j) \;   ,
\ee
where the covariance is
$$
{\rm cov}(\hat A_i, \; \hat A_j) \; = \; \frac{1}{2} \; \lgl \; \hat A_i \; \hat A_j + 
 \hat A_j \; \hat A_i  \; \rgl - 
\lgl \; \hat A_i \; \rgl \; \lgl \; \hat A_j \; \rgl .
$$
By the above definition, the covariance is symmetric:
\be
\label{213}
{\rm cov}(\hat A_i, \; \hat A_j) \; = \; {\rm cov}(\hat A_j, \; \hat A_i) \;  .
\ee
The variance (\ref{212}) can be represented as
\be
\label{214}
{\rm var}\left( \sum_i \hat A_i \right) \; = \; \sum_i {\rm var}(\hat A_i) + 
\sum_{i\neq j} \lbd_{ij} \sqrt{ {\rm var}(\hat A_i)\; {\rm var}(\hat A_j) }  \;  ,
\ee
where
\be
\label{215}
\lbd_{ij} \; = \; 
\frac{ {\rm cov}(\hat A_i,\; \hat A_j)}
{ \sqrt{{\rm var}(\hat A_i)\; {\rm var}(\hat A_j)} } \; .
\ee
Using the properties of quadratic forms \cite{Scharlau_79}, it is possible to prove
\cite{Yukalov_57,Yukalov_68} the inequality
\be
\label{216}
 |\; {\rm cov}(\hat A_i, \; \hat A_j) \; |^2 \; \leq \; 
{\rm var}(\hat A_i) \; {\rm var}(\hat A_j) \;  .
\ee
Therefore $|\lambda_{ij}| \leq 1$. Substituting relations (\ref{210}) into (\ref{214})
yields the equality
\be
\label{217}
c N^\al \; = \; \sum_i c_i N^{\al_i} + 
\sum_{i\neq j} \lbd_{ij} \; \sqrt{c_i \; c_j} \; N^{(\al_i+\al_j)/2} \; .
\ee
From here, noticing that
$$
 \sup_{i,j} \; \left( \frac{\al_i+\al_j}{2}\right) \; \leq \; \sup_i \al_i \;  ,
$$
we come to result (\ref{211}). $\square$

\vskip 2mm

{\bf Corollary}. The variance ${\rm var}{\hat A}$ of an observable quantity, behaving at 
a large number of particles $N$, according to (\ref{210}) as $N^\alpha$, is called
{\it thermodynamically normal}, provided that $\alpha \leq 1$, and is called 
{\it thermodynamically anomalous} if $\alpha > 1$. The above theorem shows that the 
variance of a composite observable is thermodynamically normal if and only if all 
partial variances are normal, and is thermodynamically anomalous if and only if at 
least one of the partial variances is anomalous. 

\vskip 2mm

The theorem tells that in multicomponent systems, where the total number-of-particle
operator is the sum 
\be
\label{218}
\hat N \; = \; \sum_i \hat N_i
\ee
of linearly independent number-of-particle operators, the overall system is stable then 
and only then when all components are stable, so that all particle fluctuations are 
thermodynamically normal, satisfying the conditions
\be
\label{219}
0 \; \leq  \frac{{\rm var}(\hat N)}{N} \; < \infty \; , \qquad 
0 \; \leq \;  \frac{{\rm var}(\hat N_i)}{N} \; < \infty \; ,
\ee
for any number of particles, including $N$ tending to infinity. 

The problem of the composite system stability can arise in the consideration of 
Bose-condensed mixtures of several species (see, e.g. \cite{Rakhimov_99}).

\section{Particle fluctuations and ensemble equivalence}

One can often meet the statement that canonical and grand canonical ensembles are not
equivalent, since particle fluctuations in these ensembles are different. However, as 
is explained above, usually the difference appears because of the incorrect use of 
these ensembles. When treating a Bose system with Bose-Einstein condensate, one has to 
remember that the necessary and sufficient condition for Bose-Einstein condensation 
is global gauge symmetry breaking. For a Bose condensed system, the use of the grand 
canonical ensemble without gauge symmetry breaking has no sense, since it does not 
represent a condensed system. And there is no any meaning in comparing not 
representative ensembles, or when at least one of them is not representative. A 
representative ensemble is the ensemble correctly representing the system of interest
\cite{Yukalov_57,Yukalov_61,Yukalov_80,Yukalov_81}, including all conditions required
for that correct description \cite{Yukalov_57,Yukalov_61,Birman_82}. Only 
representative ensembles, correctly characterizing the considered system, can be 
compared with each other \cite{Touchette_83,Touchette_84}. With regard to systems with 
Bose-Einstein condensate, both ensembles, canonical, as well as grand canonical with 
gauge symmetry breaking, are completely equivalent and equally describe all particle 
fluctuations. 

Sometimes, one thinks that for the same physical system there can exist different 
regimes, canonical, where ${\rm var}({\hat N}) \approx 0$ and grand canonical, where 
${\rm var}({\hat N}) \approx N^2$. The latter expression can be obtained from equation 
(\ref{27}) connecting the pair correlation function and particle variance, by assuming 
that the pair correlation function is uniform and equals $g(0)$, thus getting the wrong 
relation ${\rm var}({\hat N}) = N^2 (g(0) - 1)$. However, this assumption is not correct. 
The difference $g({\bf r}) - 1$ is nonzero only in a small vicinity of ${\bf r} \ra 0$, 
outside of which it becomes zero to satisfy condition (\ref{29}). The pair correlation 
function is essentially nonuniform, which follows from its definition. Thus in the HFB 
approximation it reads as
\be
\label{220}
 g(\br) \; = \; 1 + 
\frac{2}{\rho} \; [\; \rho_1(\br,0) + \sgm_1(\br,0) \; ] \; ,
\ee
which is equivalent to
\be
\label{221}
g(\br) \; = \; 1 + \frac{2}{\rho} \int ( n_k + \sgm_k) \; e^{i\bk \cdot \br} \;
\frac{d\bk}{(2\pi)^3} \;  .
\ee
The normal, $\rho_1({\bf r},0)$, and anomalous, $\sigma_1({\bf r},0)$, density 
matrices are fastly decaying functions of the distance $|{\bf r}|$. This expression 
(\ref{221}) gives the physically correct variance (\ref{148}) and compressibility 
(\ref{149}).  

The pair correlation function (\ref{221}) yields the local pair correlation
\be
\label{222}
g(0) \; = \; 1 + \frac{2}{\rho} \; ( \rho_1 + \sgm_1) \;   , 
\ee
where 
\be
\label{223}
 \rho_1 \; = \; \int n_k \; e^{i\bk \cdot\br} \; \frac{d\bk}{(2\pi)^3} \; ,
\qquad
\sgm_1 \; = \; \int \sgm_k \; e^{i\bk \cdot\br} \; \frac{d\bk}{(2\pi)^3} \;  .
\ee
For weak interactions, the local pair correlation is close to one. For example, in 
the case of particles with local interactions characterized by a scattering length 
$a_s$, the normal density $\rho_1$ and the anomalous density $\sigma_1$ are 
calculated in \cite{Yukalov_17,Yukalov_62,Yukalov_85,Yukalov_86}. For the small gas 
parameter
\be
\label{224}
 \gm \; \equiv \; a_s \; \rho^{1/3} \;  ,
\ee
the expansions are
\be
\label{225}
\frac{\rho_1}{\rho} \; \simeq \; \frac{8}{3\sqrt{\pi}} \; \gm^{3/2} + 
\frac{64}{3\pi} \; \gm^3 \; , \qquad
 \frac{\sgm_1}{\rho} \; \simeq \; \frac{8}{\sqrt{\pi}} \; \gm^{3/2} + 
\frac{64}{3\pi} \; \gm^3 \;   .
\ee
Then the local pair correlation becomes
\be
\label{226}
g(0) \; \simeq \; 1 + \frac{64}{3\pi} \; \gm^{3/2} + 
\frac{128}{3\pi} \; \gm^3 \;    .
\ee
It is evident that the pair correlation function $g({\bf r})$, that is required for 
calculating the particle variance (\ref{27}), has little to do with the local
correlation (\ref{222}). 

The pair correlation function (\ref{220}) cannot be uniform, since both, 
$\rho_1({\bf r},0)$ and $\sigma_1({\bf r},0)$ quickly diminish with the increase of
the distance $|{\bf r}|$. The momentum distribution $n_k$ and the anomalous average 
$\sigma_k$ in the integral (\ref{221}) diminish with the increasing momentum $k$.
In addition, the exponential in (\ref{221}) becomes strongly oscillating with the 
increase of $k$. This tells us that the main contribution in the integral is due to
small $k$, which suggests to approximate the sum $n_k + \sigma_k$ in the integral 
(\ref{221}) by its long-wave limit (\ref{146}). Then one gets the expression
\be
\label{227}
g(\br) \; \cong \; 1 + 
\frac{1}{\rho} \; \left( \frac{T}{m^* c^2} \; - \; 1 \right) \; \dlt(\br) \; ,
\ee
which emphasizes that the pair correlation function cannot be a constant. 

Once again, it is necessary to stress that the variance ${\rm var}(\hat N) = N^2$ 
is catastrophic, and a system with such a variance is thermodynamically unstable, 
as is explained above. If one does not forget that the use of the grand canonical 
ensemble necessarily requires the broken gauge symmetry, one never can get the 
catastrophic fluctuations that, actually, cannot exist at all. When one does not 
make mistakes in calculations, or in the experiment interpretations, one can never 
get the catastrophic particle fluctuations with the variance proportional to $N^2$. 

Particle fluctuations have also been studied employing particle-conserving ensembles,
canonical and microcanonical. In the case of a fixed number of particles $N$, by 
definition, ${\rm var}(\hat N) = 0$, hence ${\rm var}(\hat N_0) = {\rm var}(\hat N_1)$. 
Recall that for the grand canonical ensemble with gauge symmetry breaking there are no 
any condensate fluctuations, ${\rm var}(\hat N_0) \equiv 0$. This shows that the 
separation of the condensate fluctuations from the total fluctuations is ambiguous 
\cite{Idziaszek_87}. The ambiguity of separating the fluctuations of condensed and 
uncondensed particles is physically evident, as far as experimentally one observes the 
whole system in which there is no any visible separation between the components. 
Therefore, when comparing particle fluctuations in different ensembles, one, actually, 
compares the fluctuations of uncondensed particles described by ${\rm var}({\hat N}_1)$. 
In that sense, the number-conserving ensembles are equivalent to the grand canonical 
ensemble with the gauge symmetry breaking, as far as in both cases the problem is 
reduced to the consideration of fluctuations of uncondensed particles.   

For three-dimensional harmonic potentials, the ideal and weakly interacting Bose gases 
have been considered \cite{Politzer_88,Gajda_89,Kruk_90}, the results obtained in the 
canonical and microcanonical ensembles agree with the conclusions of the present paper 
and with experimental studies \cite{Kristensen_91,Christensen_92}.  

A nice analysis has been accomplished by Idziaszek \cite{Idziaszek_93} who compared 
particle fluctuations of three-dimensional ideal and interacting gases, calculated in 
the frame of the canonical as well as microcanonical ensembles for particles in a  
rectangular box and in a harmonic potential. 

For the ideal gas in a box, the microcanonical and canonical fluctuations are anomalous, 
with the variance scaling as ${\rm var}({\hat N}_1) \propto N^{4/3}$. The microcanonical 
and canonical fluctuations become equal in the thermodynamic limit. The same scaling 
exists for the grand canonical ensemble employed in the present paper. 

For the ideal gas in a harmonic potential, both fluctuations are normal, scaling as 
${\rm var}({\hat N}_1) \propto N$, but with slightly different prefactors in the 
thermodynamic limit. The scaling is in agreement with that in the grand canonical 
ensemble of the present paper.  

Interacting particles in a box, as well as in a harmonic potential, demonstrate 
canonical and microcanonical fluctuations with the scaling 
${\rm var}({\hat N}_1) \propto N$, if to stay in the range of approximation applicability, 
and scale as ${\rm var}({\hat N}_1) \propto N^{4/3}$, if to go beyond the approximation 
applicability. In the thermodynamic limit, the canonical and microcanonical variances 
always coincide. The same scaling arises in the grand canonical ensemble used in the 
present paper. 

Thus, if not to concentrate on the prefactors that can be slightly different because of
the involved approximations, from the most important point of view of the scaling all 
ensembles, grand canonical, canonical, and microcanonical, are equivalent.    
  
In the case of the ideal gas in a box, for which the variance can be treated exactly, 
the occurrence of anomalous fluctuations shows that this gas is unstable. However,
in the case of interacting gases, anomalous fluctuations, hence instability, appear     
because of the technical mistake of going beyond the frame of applicability of the 
used approximation, as is explained in Sec. 10. The anomalous fluctuations of the 
ideal gas in a box are real, signifying its instability, while the anomalous 
fluctuations of the interacting gas are spurious, caused by going beyond the 
approximation applicability. For interacting particles, the divergence in the quantity
${\rm var}({\hat N}_1)/N$ should disappear when one takes into account the attenuation 
in the spectrum of collective excitations.

One argues that the Bogolubov approximation, employed for describing particle 
fluctuations, gives reasonable results for some averages, so that why should it be 
inappicable for calculating fluctuations? It is correct that the Bogolubov-type 
approximations can often give good results, but there is no guarantee that they
can be always used outside of the region of their applicability. For example, the 
self-consistent Hatree-Fock-Bogolubov approach \cite{Yukalov_13,Yukalov_16,
Yukalov_21,Yukalov_23,Yukalov_57,Yukalov_58,Yukalov_59,Yukalov_60,Yukalov_61} does 
yield very good results for the ground-state energy of a Bose-condensed system 
\cite{Yukalov_62} and for the condensate fraction of homogeneous as well as for 
nonuniform trapped systems \cite{Yukalov_62,Yukalov_85,Yukalov_94}, rather accurately 
agreeing with Monte Carlo simulations \cite{Dubois_95,Dubois_96,Rossi_97} even for 
sufficiently strong interactions. However, this does not mean that it is always 
admissible to go beyond the approximation applicability without getting unpleasant 
surprises. Recall that the Bogolubov-type approximations (or equivalent hydrodynamic 
approximation) are of second order with respect to the field operators of uncondensed 
particles, while for calculating particle fluctuations requires to treat fourth-order
operator products.     

The other arguments that one sometimes adduces is as follows. Okay, it is possible 
to agree that the anomalous particle fluctuations do imply the system instability for 
large $N \gg 1$, as far as the compressibility becomes divergent. Of course, this is 
correct for such large statistical systems, but maybe these anomalous fluctuations 
are admissible for small statistical systems? However, this argument is of no value, 
since all statistical systems assume that $N \gg 1$. For small systems, with $N \sim 1$, 
statistical laws are not applicable at all. It would look quite strange if for some 
systems, with $N \gg 1$, the approximation would work, while for others, also with 
$N \gg 1$, it would not. It is the basis of any statistical theory that the operator 
${\hat A}$ of an observable, in order that this observable be well defined, must 
satisfy the inequality 
$$
\left| \; \frac{{\rm var}(\hat A)}{\lgl \; \hat A \; \rgl} \; \right| \; < \; \infty  
$$ 
for any size of a statistical system. A particular case is condition (\ref{31}) for 
the number-of-particle fluctuations.     
      
To summarize the present section, let us emphasize that in a given physical system
there are no different regimes with different particle fluctuations, being either
grand canonical or canonical. Particle fluctuations in any ensemble are the same,
all ensembles being equivalent with respect to the dependence on the number of 
particles in the system. The choice of an ensemble is a matter of convenience of
calculations, provided these calculations are correct. The main technical difference
is that by the formal definition of ensembles one always has one of two possibilities:

(i) {\it Grand canonical ensemble with gauge symmetry breaking}. Then, because of the
symmetry breaking, the number of condensed particles is a constant, 
$\hat N_0 = N_0 \hat 1$. There are no condensate fluctuations, so that
\be
\label{228}
{\rm var}(\hat N_0) \; = \; 0 \; , \qquad 
{\rm var}(\hat N) \; = \; {\rm var}(\hat N_1) \; .
\ee

(ii) {\it Number-preserving ensemble, canonical or microcanonical}. Then the total 
number of particles is fixed, ${\hat N} = N {\hat 1}$, so that
\be
\label{229}
 {\rm var}(\hat N) \; = \; 0 \; , \qquad 
{\rm var}(\hat N_0) \; = \; {\rm var}(\hat N_1)   .
\ee
 
{\it No matter which statistical ensemble is used, what is calculated are the 
fluctuations of uncondensed particles. And what can be measured in experiments are also
only the fluctuations of uncondensed particles.}

\section{Conclusion}

Several important problems, related to particle fluctuations in systems with 
Bose-Einstein condensate, are thoroughly studied. Unfortunately, till the present days 
there often occurs misunderstanding of the basic points in the theory of Bose-Einstein 
condensation and the resulting confusion leads to principal mistakes in the 
interpretation of experiments. Particle fluctuations in a system with Bose-Einstein 
condensate is one of such poorly understood points. The aim of the present review is
to clarify the basic points concerning particle fluctuations. Hopefully, the above
detailed explanations can persuade any thoughtful reader in the following:

\begin{enumerate}
\item
A thermodynamically stable statistical system requires that its isothermal 
compressibility be finite (Sec. 2).

\item
The isothermal compressibility is finite then and only then, when the relative 
particle variance ${\rm var}({\hat N}_1)/N$ is finite (Sec. 3).

\item 
Global gauge symmetry breaking is the necessary and sufficient condition for the
Bose-Einstein condensation (Sec. 5).

\item
The so-called ``grand canonical catastrophe" for Bose-condensed systems does not exist, 
being caused by the incorrect use of the grand canonical ensemble without gauge symmetry
breaking (Sec. 7).

\item 
In the frame of the grand canonical ensemble with gauge symmetry breaking, condensate 
fluctuations are absent at all (Sec. 7).

\item
Ideal uniform Bose-condensed gas in three dimensions is unstable having divergent
isothermal compressibility (Sec. 8).

\item
Particle interactions, irrespectively of their strength, make three-dimensional 
Bose-condensed systems stable (Sec. 10). 

\item
For a Bose-condensed system in three dimensions, particle fluctuations can become
anomalous only owing to the mistake of going beyond approximation applicability (Sec. 10).

\item
Ideal Bose-condensed gas in a rectangular trap is stable only for the spatial 
dimension $d > 4$ (Sec. 11).

\item
Ideal Bose-condensed gas in a power-law trap is stable only for the confining 
dimension $D > 2$ (Sec. 12).     

\item 
The variance of an operator that is a sum of linearly independent operators is
thermodynamically normal if and only if the variances of all partial operators are
thermodynamically normal. This implies that the particle fluctuations of a system 
composed of several species are thermodynamically normal if and only if the 
fluctuations of all components are thermodynamically normal, while if at least one
of the components exhibits thermodynamically anomalous particle fluctuations, the whole
system also has anomalous particle fluctuations (Sec. 13).

\item
Particle fluctuations in any statistical ensemble, grand canonical, canonical, or 
microcanonical, are characterized by the fluctuations of uncondensed particles (Sec. 14). 

\item 
All statistical ensembles, grand canonical, canonical, and microcanonical, are 
equivalent with respect to the scaling of particle fluctuations, hence with respect 
to system stability (Sec. 14).    

\item
Since catastrophic particle fluctuations, with the variance 
${\rm var}(\hat N) \propto N^2$, do not exist, they cannot be observed in any 
experiment, provided its interpretation is correct. 
\end{enumerate}

\section*{Acknowledgement} 

The author is grateful for discussions to V.S. Bagnato and E.P. Yukalova.

\section*{Statement}

This research did not receive any specific grant from funding agencies in the public, 
commercial, or not-for-profit sectors.

\end{document}